\documentclass{article}

\usepackage{graphicx}
\usepackage{dcolumn}

\usepackage{standalone} 
\usepackage{subfigure}  
\usepackage{tikz}
\usetikzlibrary{calc,shapes,arrows}

\usepackage{authblk}

\usepackage{mathtools}
\usepackage{amsmath,amsfonts}
\usepackage{booktabs}
\usepackage{tabularx}

\usepackage{placeins}

\graphicspath{{Figures/}}%

\usepackage[english]{babel}
\usepackage{hyperref}

\makeatletter
\def\user@resume{resume}
\def\user@intermezzo{intermezzo}
\newcounter{previousequation}
\newcounter{lastsubequation}
\newcounter{savedparentequation}
\renewenvironment{subequations}[1][]{%
      \def\user@decides{#1}%
      \setcounter{previousequation}{\value{equation}}%
      \ifx\user@decides\user@resume 
           \setcounter{equation}{\value{savedparentequation}}%
      \else  
      \ifx\user@decides\user@intermezzo
           \refstepcounter{equation}%
      \else
           \setcounter{lastsubequation}{0}%
           \refstepcounter{equation}%
      \fi\fi
      \protected@edef\theHparentequation{%
          \@ifundefined {theHequation}\theequation \theHequation}%
      \protected@edef\theparentequation{\theequation}%
      \setcounter{parentequation}{\value{equation}}%
      \ifx\user@decides\user@resume 
           \setcounter{equation}{\value{lastsubequation}}%
         \else
           \setcounter{equation}{0}%
      \fi
      \def\theequation  {\theparentequation  \alph{equation}}%
      \def\theHequation {\theHparentequation \alph{equation}}%
      \ignorespaces
}{%
  \ifx\user@decides\user@resume
       \setcounter{lastsubequation}{\value{equation}}%
       \setcounter{equation}{\value{previousequation}}%
  \else
  \ifx\user@decides\user@intermezzo
       \setcounter{equation}{\value{parentequation}}%
  \else
       \setcounter{lastsubequation}{\value{equation}}%
       \setcounter{savedparentequation}{\value{parentequation}}%
       \setcounter{equation}{\value{parentequation}}%
  \fi\fi
  \ignorespacesafterend
}
\makeatother

\newcommand{\epot}{\ensuremath{\phi} }
\newcommand{\ions}{\ensuremath{N_\mathrm{p}} }
\newcommand{\Nrif}{\ensuremath{N_\mathrm{ref}} }
\renewcommand{\v}{\ensuremath{\vec{v}} }
\newcommand{\p}{\ensuremath{\widetilde{p}} }
\newcommand{\g}{\ensuremath{\vec{g}} }
\newcommand{\E}{\ensuremath{\vec{E}} }
\newcommand{\En}{\ensuremath{E_{n}} }
\newcommand{\Eon}{\ensuremath{E_\mathrm{on}} }
\newcommand{\Erif}{\ensuremath{E_\mathrm{ref}} }
\newcommand{\T}{\ensuremath{T} }
\newcommand{\Trif}{\ensuremath{\T_\mathrm{ref}} }

\newcommand{\density}{\ensuremath{\rho} }
\newcommand{\viscosity}{\ensuremath{\nu} }

\newcommand{\specHeat}{\ensuremath{C_{V}} }
\newcommand{\thermalExpansivity}{\ensuremath{\beta_\mathrm{exp}} }
\newcommand{\heatDiff}{\ensuremath{k} }
\newcommand{\q}{\ensuremath{q} }
\newcommand{\J}{\ensuremath{\vec{j}} }
\newcommand{\Jn}{\ensuremath{j_n} }
\newcommand{\Jsat}{\ensuremath{j_\mathrm{sat}} }
\newcommand{\mobility}{\ensuremath{\mu} }
\newcommand{\diffusion}{\ensuremath{D} }
\newcommand{\eps}{\ensuremath{\varepsilon} }
\newcommand{\I}{\ensuremath{i} }
\newcommand{\Vapp}{\ensuremath{V} }
\newcommand{\Von}{\ensuremath{V_\mathrm{on}} }
\newcommand{\alphaT}{\ensuremath{\alpha_{_\mathrm{T}}} }
\newcommand{\gammaT}{\ensuremath{\gamma_{_\mathrm{T}}} }

\newcommand{\fEHD}{\ensuremath{\vec{f}_\mathrm{EHD}} }
\newcommand{\fB}{\ensuremath{\vec{f}_\mathrm{B}} }
\newcommand{\Time}{\ensuremath{t} }
\newcommand{\alphaB}{\ensuremath{\alpha} }
\newcommand{\betaB}{\ensuremath{\beta} }
\newcommand{\kappaB}{\ensuremath{\kappa} }

\newcommand{\parder}[2]{\frac{\partial #1}{\partial #2}}
\renewcommand{\quote}[1]{\textquotedblleft #1\textquotedblright}
\newcommand{\trace}[2]{\left. #1 \right|_{#2}}

\newcommand{\ABB}{{ABB Switzerland Ltd., Corporate Research, CH-5405 Baden-D\"attwil, Switzerland}}
\newcommand{\MOX}{{MOX - Dipartimento di Matematica  \quote{F. Brioschi,} Politecnico di Milano, 20133 Milano, Italy}}
\newcommand{\CEN}{{CEN - Centro Europeo di Nanomedicina, 20133 Milano, Italy}}

\newcommand{\LEA}{{Laboratory of Electromagnetics and Acoustics, Ecole Polytechnique F\'ed\'erale de Lausanne, CH-1015 Lausanne, Switzerland}}

\newcommand{\pacs}[1]{\textbf{PACS}: {#1}}
\newcommand{\keywords}[1]{\textbf{Keywords}: \textit{#1}}

\newcommand{\onlinecite}[1]{\cite{#1}}

\newcommand{\thetitle}{Multiphysics simulation of corona discharge induced ionic wind}

\usepackage{fancyhdr}
\usepackage{datetime}
\newdateformat{MonthYY}{\monthname[\THEMONTH]\ \THEYEAR}
\begin{document}

\title{\thetitle}

\author[1,2]{Davide Cagnoni\thanks{Corresponding author: davide.cagnoni@polimi.it}}
\author[2]{Francesco Agostini}%
\author[2]{Thomas Christen}%
\author[1,3]{Carlo de Falco}%
\author[1]{Nicola Parolini}%
\author[2,4]{Ivica Stevanovi\'c}%
\affil[1]{\MOX}
\affil[2]{\ABB}
\affil[3]{\CEN}
\affil[4]{\LEA}%
\date{\today}

\maketitle             

\begin{abstract}
Ionic wind devices or electrostatic fluid accelerators are becoming of increasing interest as tools for thermal management, in particular for semiconductor devices. In this work, we present a numerical model for predicting the performance of such devices, whose main benefit is the ability to accurately predict the amount of charge injected at the corona electrode.
Our multiphysics numerical model consists of a highly nonlinear strongly coupled set of PDEs including the Navier-Stokes equations for fluid flow, Poisson's equation for electrostatic potential, charge continuity and heat transfer equations. To solve this system we employ a staggered solution algorithm that generalizes Gummel's algorithm for charge transport in semiconductors.
Predictions of our simulations are validated by comparison with experimental measurements and are shown to closely match. Finally, our simulation tool is used to estimate the effectiveness of the design of an electrohydrodynamic cooling apparatus for power electronics applications.



\end{abstract}

\FloatBarrier
\section{Introduction and motivation}

Cooling of electric and electronic devices is a continuous challenge for researchers and engineers. Power electronics trends indicate a continuous increase of power densities and a shrink of component dimensions. These conditions make the thermal management a pillar to guarantee a safe, reliable and affordable operation of electronic components where suitable cooling schemes must be applied. Forced convection air cooling is probably the oldest and still one of the most used approaches for electronic systems cooling. Usually, forced convection is driven by a fan but, for some applications as, for example, the cooling of hot spots or enclosure-contained devices, alternative methods based on \emph{Electro-Hydrodynamic} (EHD) forces have been recently studied and exploited.
A representative example of such methods is that of ionic wind induced by a so called \emph{corona discharge}. 

\begin{figure}[tb]
   \centering
   \subfigure[\ Overall scheme of positive corona induced ionic wind.\label{fig:ionic:wind}]{\includegraphics[width=.48\textwidth]{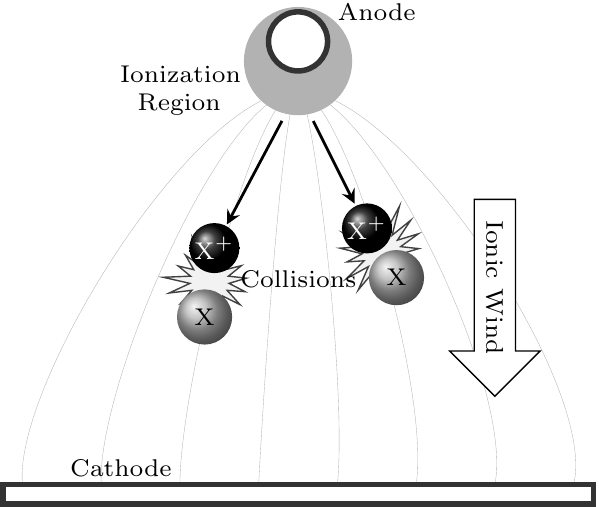}}
   \subfigure[\ Zoom on the ionization region to show avalanche charge multiplication.\label{fig:ionic:wind:zoom}]   {\includegraphics[width=.48\textwidth]{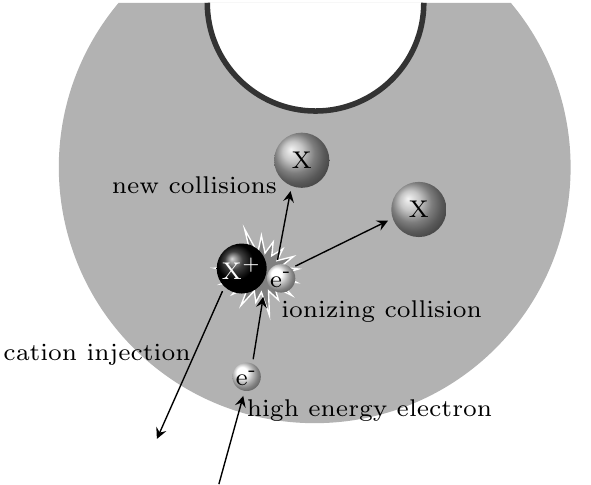}}
   \caption{Schematic representation of positive wire-to-plane corona discharge-induced ionic wind. In ambient air, $\mathrm{X}$ represents primarily $\mathrm{O}_2$ or $\mathrm{N}_2$ molecules, and the dominant ionization reactions are of the type $\mathrm{e}^- + \mathrm{X} \rightleftharpoons 2\mathrm{e}^- + \mathrm{X}^+$.\cite{zhang-04}
     \label{fig:ionic:wind:and:zoom}} 
\end{figure}

Figure~\ref{fig:ionic:wind:and:zoom} schematically illustrates the phenomenon of corona discharge occurring between two electrodes in air. 
The gas ions formed in the discharge are accelerated by the electric field and exchange momentum with neutral fluid molecules, initiating a drag of the bulk fluid which is referred to as ionic wind. The choice of a positive corona is favorable in industrial applications as it leads to significantly reduced ozone production, and increased durability of the metal electrodes in comparison to negative corona devices.\cite{castle-tiga-69} 
Therefore, in this study, we focus on the case of DC positive corona wind, where the applied voltage at the electrodes is stationary, gas ionization occurs at the anode and charge carriers are mainly $\mathrm{O}_{2}^{+}$ ions, as described in Fig.~\ref{fig:ionic:wind}. 

Both experimental and numerical studies of EHD phenomena have been presented in recent literature.
For example, Adamiak and others \cite{adamiak-04,adamiak-04b,sattari-adamiak-10,zhao-esjc-09} studied the DC and pulsed corona discharge between a needle and a plate collector, using different numerical methods (FEM, BEM, FCT etc.) for the approximation of each equation in the PDE system; Ahmedou and Havet \cite{ahmedou-je-09,ahmedou-09-b,ahmedou-09-2} used a commercial FEM software to investigate the effect of EHD on turbulent flows;
Moreau and Touchard,\cite{moreau-je-08}  Huang and others,\cite{Huang20091789} and Kim and others\cite{kim-je-10} experimentally studied different EHD devices designed for cooling or air pumping purpose; Chang, Tsubone and others \cite{chang-jpd-07,tsubone-je-08,chang-2009} made extensive experimental study of the forced airflow and the corona discharge in a converging duct; Jewell-Larsen and others \cite{karpov-comsol-05,jewell-larsen-06,larsen-tdei-06,larsen-jms-07,larsen-esa-08,larsen-tdei-08,larsen-semitherm-09} and Go and others \cite{go-06,go-jap-07,fisher-07-patent,go-ijhmt-08,go09} conducted both experimental and numerical studies  aimed at designing and applying ionic wind cooling devices to thermal management of electronic devices.

In this paper, we use a numerical approach based on a multiphysics mathematical model that accounts for all relevant electrostatic, fluid, and thermal aspects of the phenomena being considered. Particular attention is devoted to correctly modeling the relation between the electric field at the anode and the amount of charge injected from the anode corona into the neutral gas region. 
The accuracy of such relation is crucial for increasing the predictive capability of numerical simulations. Here, we present a novel approach for modeling charge injection, which is based on enforcing Kaptsov's hypothesis\cite{kaptsov-47} and is shown to provide good simulation accuracy using few free model parameters.
Our approach to the charge injection modeling is compared to those existing in the literature on a set of benchmark device geometries for which experimental data are available.

\FloatBarrier
\section {Governing equations} \label{sec:gov:eq}

\begin{figure}[tb]
\centering
\includegraphics[width=.65\textwidth]{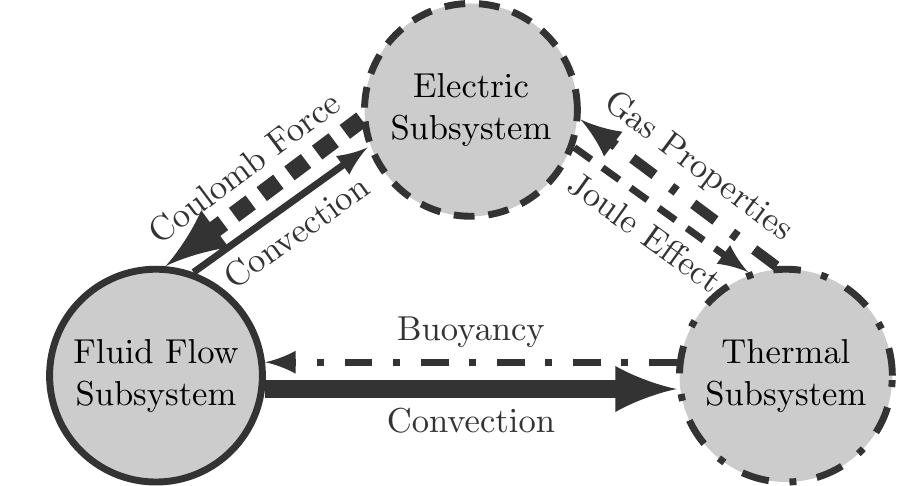}
\caption{Relations between the variables in the EHD system, with arrows pointing to an influenced subsystems from the influencing one. Thicker arrows indicate stronger interactions, while thinner ones indicate minor influence. The chart is adapted from Ref.~\onlinecite{yabe-aiaa-78}. \label{fig:equation:relations}}
\end{figure}

\begin{figure}[tb]
\centering
\includegraphics[width=.65\textwidth]{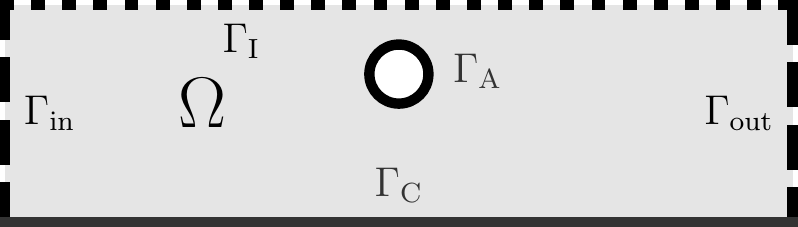}
\caption{\label{fig:example:domain} Example domain where all the five possible kinds of boundary are depicted.}
\end{figure}

Modeling of EHD systems requires accounting for a number of interplaying phenomena of different physical nature. Figure~\ref{fig:equation:relations} summarizes such phenomena and their interactions: electric current due to drifting ions generates bulk fluid flow which, in turn, contributes to ion drift; thermal energy is transported by the flowing fluid while, at the same time, temperature gradients give rise to buoyancy forces; finally electric conduction properties of the gas are influenced by temperature, while electric currents act as heat sources via Joule effect. 
 
The system of partial differential equations governing the behavior of each subsystem is introduced below together with most of the constitutive relations for the system coefficients. The PDEs described below are set in an open bounded domain $\Omega$ whose typical geometry is shown in Fig.~\ref{fig:example:domain}; the domain $\Omega$ represents the region in space occupied by bulk neutral fluid and drifting positive ions. In our model, the thickness of the ionization layer around the anode is considered to be negligible with respect to the length scale of the overall system. Such region is therefore represented as a portion of the boundary, denoted as $\Gamma_\mathrm{A}$ in Fig.~\ref{fig:example:domain}, and the process of ion injection is modeled by enforcing a suitable set of boundary conditions on $\Gamma_\mathrm{A}$. Existing and new models for such boundary conditions are discussed in Section~\ref{sec:bcs}.

Unipolar (positive) electrical discharge in fluid is described by Poisson's equation
\begin{subequations}\label{eq:EHD:system}
\begin{equation}
 \nabla \cdot (\eps\E) = -\nabla \cdot  (\eps\nabla\epot ) = \q\ions,
 \label{eq:poisson}
\end{equation}
coupled with current continuity equation
\begin{equation}
 \parder{\q\ions}{\Time} + \nabla\cdot\J = 0,
 \label{eq:cons:charge}
\end{equation}
\end{subequations}
where \eps is the electrical permittivity, \E the electric field, \epot the electric potential, \q the elementary (proton) charge, \ions the number density of (positive) ions. 
The current density \J is given by the sum of three contributions: drift due to electric field, advection, and diffusion:
\begin{equation}
\J=\q\ions\left(\mobility\E+\v\right) -\q\diffusion\nabla\ions,
\label{eq:current:density:mobility}
\end{equation}
\mobility being the ion mobility in the fluid and \v the fluid velocity field. The diffusion rate \diffusion is related to mobility and temperature \T through Einstein's relation
\begin{equation}
\diffusion = \mobility k_\mathrm{B} \T \q^{-1}, 
\end{equation}
where $k_\mathrm{B}$ is Boltzmann's constant. 
The flow of incompressible Newtonian fluids is described by the Navier-Stokes equations, which represent the conservation of momentum and mass density:
\begin{subequations}[resume]
\begin{equation}
\label{eq:navier:stokes:incomp}
\begin{cases}
\displaystyle \parder{\v}{t} + (\v\cdot \nabla) \v =\viscosity\Delta\v - \nabla {\p} +  \frac{\fEHD + \fB }{\density}, &
\\
\displaystyle \nabla \cdot \v =0,
&
\end{cases}
\end{equation}
\end{subequations}
where \viscosity is the kinematic viscosity, \p is the modified (non-hydrostatic) pressure $p\density^{-1}-\g\cdot\vec{x}$, \density being the gas mass density and \g the gravity acceleration. The volume force term on the right hand side of the first equation of~\eqref{eq:navier:stokes:incomp} consists of the sum of electrohydrodynamic force \fEHD and buoyancy force $\fB$. As we consider single-phase flows with limited temperature gradients, \fEHD can be expressed as:\cite{stratton-41,jones-aht-79,yu-99}
\begin{equation}
 \fEHD = \q \ions \E.
\end{equation}
For buoyancy force, due again to the limited temperature gradients, the Boussinesq approximation can be adopted:
\begin{equation}
 \fB = \g[\density(\T)-\density] = \g\left[\density{\thermalExpansivity(\Trif)} (\T-\Trif)\right],
\end{equation}
where \thermalExpansivity is the thermal expansion coefficient, and the dependence of the gas density $\density(\T)$ on temperature \T is linearized around a certain reference temperature $\Trif$, at which the reference density \density is taken.
Finally, the temperature equation, which describes heat transfer, reads
\begin{subequations}[resume]
\begin{equation}
 \parder{\T}{t} + \v\cdot\nabla\T - \frac{\heatDiff}{\density\specHeat}\Delta\T = \frac{\dot{Q}}{\density\specHeat},
 \label{eq:temperature}
\end{equation} 
\end{subequations}
where \heatDiff is the heat conductivity and \specHeat the mass specific heat.
The thermal power production $\dot{Q}$ on the right hand side of \eqref{eq:temperature} can be expressed as a balance of terms accounting for the Joule heating caused by the current density \J and the mechanical power provided by the EHD force $\fEHD$:
\begin{equation}
\dot{Q} = \J\cdot\E - \v\cdot\fEHD = (\q\ions\mobility\E - \q\diffusion\nabla\ions)\cdot\E.
\end{equation}
In addition to the volume thermal energy generation pertaining to $\dot{Q}$, thermal energy is also generally exchanged with an external body; it is worth pointing out that, in general, the contribution of the injected energy through the system boundary usually outweighs the volume power production $\dot{Q}$, for the small electric currents flowing in EHD systems. 

The coupled system of PDEs~\mbox{\eqref{eq:poisson}-\eqref{eq:temperature}} presented in this section needs to be completed by a suitable set of initial and boundary conditions. 
An example of computational domain $\Omega$ is shown in Fig.~\ref{fig:example:domain}; the domain boundary $\partial\Omega$ is partitioned into five different subregions \mbox{$\partial\Omega=\Gamma_\mathrm{in}\cup\Gamma_\mathrm{out}  \cup\Gamma_\mathrm{I}\cup\Gamma_\mathrm{C}\cup\Gamma_\mathrm{A}$} on which different boundary conditions are enforced. Initial conditions, which are to be set for ion density, velocity, and temperature, are chosen as uniform fields, with values based on the expected \quote{device off} state. 

The fluid inlet is represented by the boundary region $\Gamma_\mathrm{in}$, where Dirichlet conditions are enforced for the velocity \v and the temperature $\T$,  and homogeneous Neumann condition is enforced on $\p$. Since the inlet is supposed to be far from the electrodes, and thus from the region where major electrical phenomena are localized, the electrical variables are also considered to have vanishing gradients along the outward normal direction $\vec{n}$ on the boundary $\partial\Omega$. 
At the fluid outlet $\Gamma_\mathrm{out}$, we require the normal component of the fluid stress tensor and of the temperature, charge density, and electric potential gradients to vanish, representing again a region which is far from the major phenomena in the system.

The boundary region denoted as $\Gamma_\mathrm{I}$ represents an electrically insulating wall and both drift and diffusion current densities ($\q\ions(\mobility\E+\v)$ and $-\q\diffusion\nabla\ions$, respectively) are supposed to independently vanish. Since $\Gamma_\mathrm{I}$ is also a solid wall, the condition of non-penetration $\v\cdot\vec{n}=0$, which we impose on the fluid flow, allows for the drift current to vanish if $\E\cdot\vec{n}=0$. Diffusion currents are instead damped by the homogeneous Neumann condition for ion density $\nabla\ions\cdot\vec{n} = 0$. Additionally, fluid flow is subject to a non-slip condition $\|\v-(\v\cdot\vec{n})\vec{n}\| = \nabla\p\cdot\vec{n} = 0$. Temperature can either assume an imposed value, or satisfy an imposed thermal energy flux trough the wall surface, depending on the situation at hand.

Finally, the regions $\Gamma_\mathrm{C}$ and $\Gamma_\mathrm{A}$ represent the cathode and anode contacts, respectively.
At both electrodes, we enforce Dirichlet condition for the electrostatic potential and no-slip, no-penetration conditions for the fluid flow.
The cathode $\Gamma_\mathrm{C}$ often coincides with the surface to be cooled, in which case we may impose either fixed heat flux through the surface, or fixed temperature, as we do on $\Gamma_\mathrm{I}$. 

With regard to the ion density, homogeneous Neumann condition is enforced on $\Gamma_\mathrm{C}$. Physically, this means that the only current allowed through the cathode is due to ion drift: since mass is not allowed to cross the boundary, though, this results in imposing each one of the positive ions hitting the cathode to recombine with an extracted electron.
Boundary conditions for ion density on the anode are instead more complicated, and Section~\ref{sec:bcs} is entirely devoted to the derivation and comparison of different models for such boundary conditions.

\FloatBarrier
\section{Modeling of charge injection}\label{sec:bcs}

To trigger the corona discharge, the voltage drop between anode and cathode must exceed a threshold (or \emph{onset}) value which we denote by $\Von$, while the corresponding magnitude of the electric field at the anode is denoted by $\Eon$. 
The generally accepted Kaptsov's hypothesis\cite{kaptsov-47} states that free charge, emitted by the corona for voltages higher than $\Von$, causes a shielding of the anode that results in \quote{clamping} of the anode electric field at the onset value $\Eon$.  While $\Von$ depends very strongly on the whole device geometry, experimental evidence indicates that \Eon is strictly correlated with the curvature radius of the anode contact.\cite{peek-29}

At the microscale, corona discharge is generated by the impact ionization of gas molecules and avalanche multiplication of electrons. According to the avalanche model first developed by Townsend \cite{townsend-00,townsend-01}, cations are generated in an area characterized as the locus of points $\vec{x}\in\Omega$ such that:
\begin{equation}\label{eq:exponential:townsend:treshold}
\gammaT\exp{\left(\int_{L(\vec{x})}\alphaT(\vec{r})\cdot\text{d}\vec{r}\right)} \geq 1,
\end{equation}
where \gammaT and \alphaT are parameters depending on the applied electric field, the pressure, and the chemical composition of the gas and the electrodes, whereas $L(\vec{x})$ is the trajectory of a negatively charged particle, which leaves from the cathode and drifts to $\vec{x}$ due to the force exerted on it by the electric field.

Although not of much practical interest when space charge is not negligible, relation~\eqref{eq:exponential:townsend:treshold} provides a rough estimate for the thickness of the ionization region, where the gas can effectively be considered to be in plasma state. Such thickness depends on the geometry of the anode as well as on the gas pressure and on the electric field; in corona discharge regime, it is so small in comparison to the \mbox{length-scale} of the neutral fluid region, that it makes sense to adopt a \emph{lumped} model for the ionization region and to represent it as a portion of the anode surface. Under such approximation, the only charge carriers within the bulk fluid region are cations.\cite{aliat-jpdap-09}

In this section we discuss several possible options for modeling the rate at which such cations are injected into the bulk fluid region. In order to ease the comparison of the different models, we will express all of them in the common form of a Robin-type boundary condition for equation~\eqref{eq:cons:charge}:
\begin{equation}\label{eq:standard:bc}
 \alphaB \trace{\ions}{\Gamma_\mathrm{A}} + 
 \betaB \trace{\partial_{\vec{n}}\ions}{\Gamma_\mathrm{A}} = \kappaB,
\end{equation}
where $\partial_{\vec{n}}\ions = \nabla\ions\cdot\vec{n}$ is the component of the ion density gradient normal to $\Gamma_{\mathrm{A}}$. The condition~\eqref{eq:standard:bc} will be in general nonlinear as we allow the coefficients $\alphaB$, $\betaB$, and \kappaB to depend locally on the normal component $\En = \vec{n}\cdot\E|_{\Gamma_\mathrm{A}}$ and on the density of ions $\trace{\ions}{\Gamma_\mathrm{A}}$.

The most common approach used in numerical studies of positive corona discharge that appeared in the literature\cite{larsen-tdei-08,mazumder-11a,mazumder-11b,zhang-lai-10} consists in imposing the current at the anode to be equal to the experimentally measured value $\I_\text{m}$. This leads to the following choice of parameters in~\eqref{eq:standard:bc}
\begin{equation}\label{eq:bc:uniform}
 \begin{cases}
    \alphaB_1 \trace{\ions}{\Gamma_\mathrm{A}} + \betaB_1 \trace{\partial_{\vec{n}}\ions}{\Gamma_\mathrm{A}} = \kappaB_1, \\[3mm]
    \alphaB_1=-\q\mobility\En,\quad \betaB_1 = \q\diffusion,\quad \kappaB_1 = \I_\text{m}/s .
 \end{cases}
\end{equation}
Notice that~\eqref{eq:bc:uniform} is based on the additional assumption that the component of the ion current density \mbox{$\Jn={\vec{n}\cdot\J}|_{\Gamma_{\mathrm{A}}}$} normal to the contact be uniformly distributed along $\Gamma_{\mathrm{A}}$ (hence, we will hereafter refer to this model as \emph{uniform}).
This latter assumption, together with the fact that knowledge of a measured value of the current corresponding to each value of the applied bias is required, strongly limits the ability of simulations based on~\eqref{eq:bc:uniform} to provide useful information about the impact of the anode contact geometry on device performance.

One possible approach to overcome the drawbacks of~\eqref{eq:bc:uniform} is to enforce a pointwise relation between \Jn and the normal component of the electric field on $\Gamma_\mathrm{A}$. Such relation, as proposed in Ref.~\onlinecite{christen-jes-07}, accounts for a balance between different contributions that make up the ionic current at the microscale:
\begin{equation}\label{eq:bc:sccc}
 \Jn = w\ions - \Jsat H(\En-\Eon),
\end{equation}
where $H(x)$ denotes Heaviside's step function. The parameters appearing in \eqref{eq:bc:sccc} are the maximum allowed current density $\Jsat$, the threshold field $\Eon$, and the proportionality constant $w$ (which has dimensions of a velocity times an electric charge) between the \emph{backscattering} current and the amount of ions accumulated in the \textit{space charge} region at the anode. This model will be hereon denoted \emph{SCCC}, as in \quote{space charge controlled current}. Using~\eqref{eq:bc:sccc} to determine the coefficients of the general expression~\eqref{eq:standard:bc} leads to
\begin{equation}\label{eq:sccc:standard:form}
  \begin{cases}
     \alphaB_2 \trace{\ions}{\Gamma_{\mathrm{A}}} + \betaB_2 \trace{\partial_{\vec{n}}\ions}{\Gamma_\mathrm{A}} = \kappaB_2, \\[3mm]
     \alphaB_2=w-\q\mobility\En,\quad \betaB_2 = \q\diffusion,\quad \kappaB_2 = \Jsat  H(\En-\Eon).
  \end{cases}
\end{equation}
While this model does not require prior knowledge of the current density, thus apparently solving the main issue of model \eqref{eq:bc:uniform}, the quality of its predictions depends critically on the correct choice of its parameters \Jsat and $w$ and appears to be, for some relevant practical situations, quite poor if these parameters are given bias-independent values.

An alternative approach consists of selecting the coefficients of~\eqref{eq:standard:bc} in such a way as to enforce, pointwise on $\Gamma_{\mathrm{A}}$ the negative feedback relation between normal  electric field and space charge that is at the basis of Kaptsov's hypotesis. This can be done, for example, by defining the following model:
\begin{equation}\label{eq:diode:standard:form}
\begin{cases}
\alphaB_3 \trace{\ions}{\Gamma_{\mathrm{A}}} + \betaB_3 \trace{\partial_{\vec{n}}\ions}{\Gamma_\mathrm{A}} = \kappaB_3 , \\
\alphaB_3 = \q\mobility\Eon, \quad \betaB_3 =0, \quad \kappaB_3 = \q\mobility\En \trace{\ions}{\Gamma_\mathrm{A}} .
\end{cases}
\end{equation}
Model \eqref{eq:diode:standard:form} has only one parameter, the onset field $\Eon$, whose typical magnitude can be, at least roughly, estimated by means of correlations available in literature.\cite{peek-29} 
On the other hand, \eqref{eq:diode:standard:form} presents a further nonlinearity in comparison to \eqref{eq:bc:uniform} and \eqref{eq:sccc:standard:form}, as $\kappaB_3$ depends on $\ions$, thus its implementation requires a suitable linearization approach. Since in this study we are mainly interested in the stationary regime device performance, we adopt the simplest approach and evaluate $\kappaB_3$ in \eqref{eq:diode:standard:form} using the latest computed value of \ions. This approach will be shown in numerical examples of Section~\ref{sec:benchmark} to be very effective in terms of accuracy of the simulation, but to also highly impact the computational time required for the simulated current to reach its regime value. This model was named \emph{ideal diode}, since it allows arbitrary currents over the threshold, and no current under the threshold.

The alternative method of solving the nonlinearity adopted, \textit{e.g.}, in Ref.~\onlinecite{adamiak-04} or Refs.~\onlinecite{jewell-larsen-06,karpov-comsol-05} does not seem to reduce such numerical problems. We are therefore lead to consider yet one more type of boundary condition at the anode, where part of the predictive accuracy of \eqref{eq:diode:standard:form} is traded off to achieve better numerical efficiency. This latter model is expressed by the following choice of the boundary condition coefficients:

 \begin{equation}\label{eq:real:diode:standard:form}
\begin{cases}
\alphaB_4\ions + \betaB_4 \partial_{\vec{n}}\ions = \kappaB_4, \\
\alphaB_4 = \q\mobility\Eon , \quad \betaB_4 =0 ,\quad 
\kappaB_4 = \q\mobility\Eon \Nrif \exp \left(\frac{\En-\Eon}{\Erif}\right),
\end{cases}
\end{equation}
where \Erif is a reference electric field and $\Nrif$ is a reference cation density. It can be easily verified that the set of points in the \mbox{$\ions$-$\En$} plane that satisfy~\eqref{eq:real:diode:standard:form} reduces to the set satisfying~\eqref{eq:diode:standard:form} as $\Erif\rightarrow 0$; in such sense, an interpretation of this model as a smoother version of the \emph{ideal diode} model is possible; to highlight the analogy with \eqref{eq:diode:standard:form}, thus, this model was named \emph{exponential diode}.

A summary of the kinds of boundary conditions considered in this paper is presented in Table~\ref{tab:bc:coeffs} where, for each condition, the corresponding models for the coefficients $\alphaB$, \betaB and \kappaB is reported. 

\begin{table}[tb]
\caption{\label{tab:bc:coeffs}Summary of the coefficients for the four boundary models presented} 
\centering
\begin{tabularx}{\textwidth}{X c c c c}
\toprule
Model name	\ &Equation\ & \alphaB 		\ & \betaB 		\ & \kappaB \\
\midrule 
\hline \\
Uniform 	\ &	\eqref{eq:bc:uniform}\ & $-\q\mobility\En$ 	\ & $\q\diffusion$ 	\ & ${\I_\text{exp}}/{s}$ \\\\
\midrule 
SCCC 		\ &	\eqref{eq:sccc:standard:form}\ & $w-\q\mobility\En$ 	\ & $\q\diffusion$ 	\ & $\Jsat H\left(\En - \Eon\right)$ \\\\
\midrule 
Ideal diode 	\ &	\eqref{eq:diode:standard:form}\ & $\q\mobility\Eon$ 			\ & $0$	 	\ & $\q\mobility\ions\En$ \\\\
\midrule 
Exponential diode 	\ &	\eqref{eq:real:diode:standard:form}\ & $\q\mobility\Eon$ 			\ & $0$ 		\ & $\q\mobility\Eon\Nrif\exp\left(\tfrac{\En-\Eon}{\Erif}\right)$ \\
\bottomrule
\end{tabularx}
\end{table}

\FloatBarrier
\section{Decoupled iterative solution algorithm}\label{sec:algo}

\begin{figure}[tb]
\includegraphics[width=.85\textwidth]{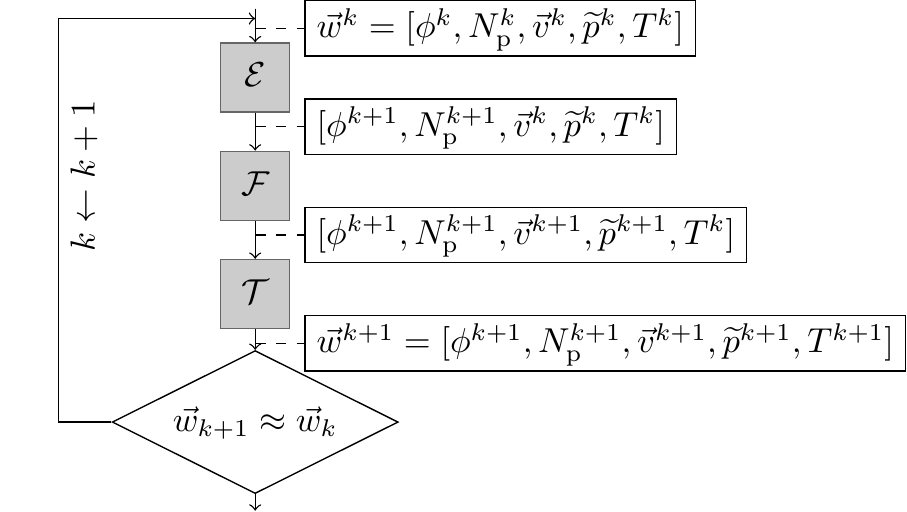}
\caption{\label{fig:fixed:point:map} Block diagram representing the composite fixed point iteration used to solve the system \eqref{eq:poisson:bc:semidiscrete}--\eqref{eq:energyeq:bc:semidiscrete}.}
\end{figure}

The algorithm we developed for the solution of system~\mbox{\eqref{eq:poisson}-\eqref{eq:temperature}} is constructed by analogy with iterative algorithms used for the solution of similar systems of coupled PDEs that arise in modeling of semiconductor devices by drift-diffusion or hydrodynamic models~\cite{Ballestra2004320,Brezzi2005317,deFalco20071729,deFalco2005533,jerome96,selberherr-84} or in  electrochemical models for ionic transport in biological systems.\cite{Chini2006,Jerome200810,Jerome2009}
The algorithm is constructed by a sequence of four steps:
\begin{enumerate}
\item time-semidiscretization by means of Rothe's method is performed to reduce the initial/boundary value problem~\mbox{\eqref{eq:poisson}-\eqref{eq:temperature}} to a sequence of boundary value problems, where only derivatives with respect to the spatial coordinates appear;
\item the sub-problems composing the whole system are decoupled and a strategy to iterate among them in order to achieve self consistency is chosen;
\item as the decoupled sub-problems are still nonlinear, inner iterations need be defined to solve them;
\item finally, as the initial problem has been reduced to a set of scalar linear problems, a proper spatial 
discretization scheme is chosen to solve them numerically. 
\end{enumerate}

We choose the Backward Euler scheme for \mbox{time-discretization}, as we are mainly interested in capturing steady state behavior rather than accurately describing transient system dynamics, and thus we favor stability over high order accuracy. 
For the sake of convenience we summarize below the full system~\eqref{eq:poisson}-\eqref{eq:temperature} as it appears after applying           \mbox{time-discretization} and enforcing the boundary conditions discussed above.

\begin{subequations}\label{eq:semidiscrete:formulation}\begin{align}
\shortintertext{\textit{Poisson equation} }
& \begin{cases}\displaystyle{\label{eq:poisson:bc:semidiscrete}}\tag{\ref*{eq:poisson}$^\prime$}
     -\nabla \cdot (\eps\nabla\epot) = \q\ions & \text{ on }\Omega\\
     \epot = \Vapp_\text{A} &\text{ on }\Gamma_\text{A}\\
     \epot = 0 &\text{ on }\Gamma_\text{C}\\
     \partial_{\vec{n}}\epot = 0 &\text{ on }\Gamma_\text{I}\cup\Gamma_\text{in}\cup\Gamma_\text{out}
  \end{cases}
\shortintertext{\textit{(Time-discretized) Current continuity equation} }
& \begin{cases}\displaystyle{\label{eq:charge:conserv:bc:semidiscrete}}\tag{\ref*{eq:cons:charge}$^\prime$}
      \frac{\q(\ions-\ions^\text{old})}{\delta\Time} + \nabla\cdot\left(-\diffusion\q\nabla\ions +(\v-\mobility\nabla\epot)\q\ions\right) = 0 & \text{ on }\Omega \\
      \alphaB\ions + \betaB\partial_{\vec{n}}\ions = \kappaB  &\text{ on }\Gamma_{\text{A}} \\
      \partial_{\vec{n}}\ions = 0 &\text{ on }(\partial\Omega\setminus\overline{\Gamma}_\text{A})
  \end{cases}
\shortintertext{\textit{(Time-discretized) Navier-Stokes equations} }
 & \begin{cases}\displaystyle{\label{eq:navier-stokes:bc:semidiscrete}}\tag{\ref*{eq:navier:stokes:incomp}$^\prime$}
      \frac{\v-\v^\text{old}}{\delta\Time}-\nabla\cdot(\viscosity\nabla\v) + (\v\cdot\nabla)\v -\nabla\p = \frac{\fEHD + \vec{f}_\text{b}}{\density}  & \text{ on }\Omega\\
      \nabla \cdot \v = 0 &\text{ on }\Omega\\
      \v= 0  &\text{ on }\Gamma_\text{A}\cup\Gamma_\text{C}\cup\Gamma_\text{I}\\
      \v= \v_\text{in} &\text{ on }\Gamma_\text{in}\\
      -\viscosity\partial_{\vec{n}}\v + \p\vec{n} = 0 &\text{ on }\Gamma_\text{out}
  \end{cases}
 \shortintertext{\textit{(Time-discretized) Heat equation} }
 & \begin{cases}\displaystyle{\label{eq:energyeq:bc:semidiscrete}}\tag{\ref*{eq:temperature}$^\prime$}
      \frac{\density\specHeat(\T-\T^\text{old})}{\delta\Time} + \nabla\cdot(-\heatDiff\nabla\T + \v\density\specHeat\T) =   (\mobility\E\q\ions - \diffusion\q\nabla\ions)\cdot\E   & \text{ on }\Omega\\
      \heatDiff\partial_{\vec{n}}\T = e_\text{in} &\text{ on }\Gamma_\text{A}\cup\Gamma_\text{C}\cup\Gamma_\text{I}\\
      \T= \T_\text{in} &\text{ on }\Gamma_\text{in}\\
      \heatDiff\partial_{\vec{n}}\T = 0 &\text{ on }\Gamma_\text{out}
  \end{cases}
\end{align}\end{subequations}
The outer iteration strategy for decoupling system~\mbox{\eqref{eq:poisson:bc:semidiscrete}-\eqref{eq:energyeq:bc:semidiscrete}} is graphically represented in Fig.~\ref{fig:fixed:point:map}.
The equations are subdivided into three blocks representing the electrical, fluid and thermal subsystems, respectively. In Figure~\ref{fig:fixed:point:map} each subsystem is identified in terms of its {\it solution map}, namely $\mathcal{E}$ for the electrical subsystem~\mbox{\eqref{eq:poisson:bc:semidiscrete}-\eqref{eq:charge:conserv:bc:semidiscrete}}, $\mathcal{F}$ for the fluid subsystem~\eqref{eq:navier-stokes:bc:semidiscrete} and $\mathcal{T}$ for the thermal subsystem~\eqref{eq:energyeq:bc:semidiscrete}.
Each of such maps operates on a subset of the components of the complete system  state vector $\vec{w} = [\epot, \ions, \v, \p, \T]$ and iteration is performed by applying the fixed point map $\mathcal{M} = \mathcal{T} \circ \mathcal{F} \circ \mathcal{E}$ until the prescribed tolerance is achieved.

The main advantage of decoupling the system according to the physics as outlined above is that each subproblem can then be treated following a specifically tailored approach, which is known to be the most appropriate in its respective field.
In particular, the map $\mathcal{E}$ is based on the well-known Gummel-map strategy widely used in computational electronics,\cite{selberherr-84,Brezzi2005317,deFalco20071729,deFalco-09} the map $\mathcal{F}$ is composed of an incompressibility-enforcing iteration based on the standard PISO scheme,\cite{issa86} well established for the solution of incompressible Navier-Stokes equations.
Finally, the map $\mathcal{T}$ represents the solution of temperature equation, which is treated as a linear equation, neglecting the gas coefficients variations. 

The results presented in the next section have been obtained using the finite volume method (FVM) for space discretization. A custom solver has been implemented within the C++ library OpenFOAM.\cite{openfoam} However, the algorithm presented in this section is very general and could be extended to different discretization 
methods. 
  
\FloatBarrier
\section{Model validation} \label{sec:benchmark}

In this section, three different test geometries are presented, and the results obtained in our simulations are compared to experimental and numerical data. The simulations were obtained with the help of the library \texttt{swak4Foam} \cite{swak4Foam} for the implementation of the boundary conditions, while the domain meshes were produced with \texttt{gmsh} \cite{gmsh}.

\subsection{Open wire to wall-embedded collecting electrode arrangement} \label{subsec:go}

\begin{figure}[b]
\centering
 \includegraphics[width=.65\textwidth]{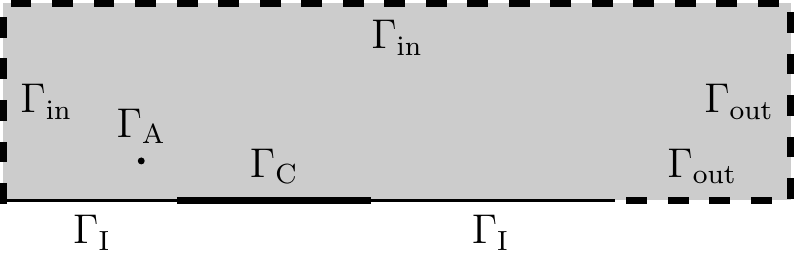}
 \caption{Scheme of the computational domain geometry for the device with open wire to wall-embedded collecting electrode arrangement discussed in Subsection~\ref{subsec:go}. \label{fig:Go:geometry}}
\end{figure}
\begin{figure}[t!] 
   \centering
   \subfigure[\ Ion number density distribution ($\mathrm{m}^{-3}$) in a device region near the electrodes. The ticks on the right show the length scale, each tick is $1\,\mathrm{mm}$.\label{fig:go:el:1}]
   {\includegraphics[width=.48\textwidth]{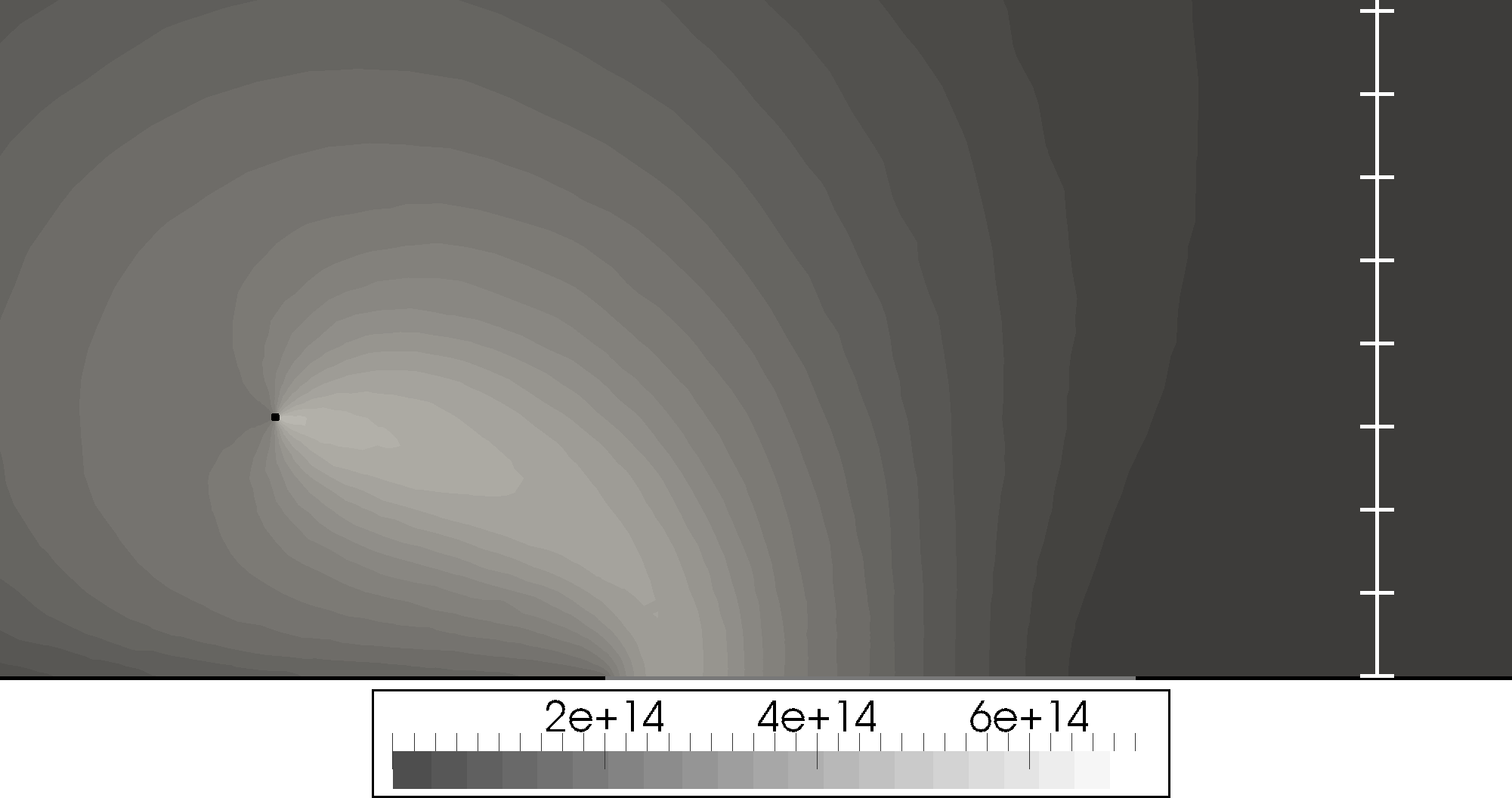}}
   \subfigure[\ Electric field lines-of-force (grey) and electric potential isolines (black) in a device region near the electrodes. The scale is the same as in Fig.~\ref{fig:go:el:1}.\label{fig:go:el:2}]{\includegraphics[width=.48\textwidth]{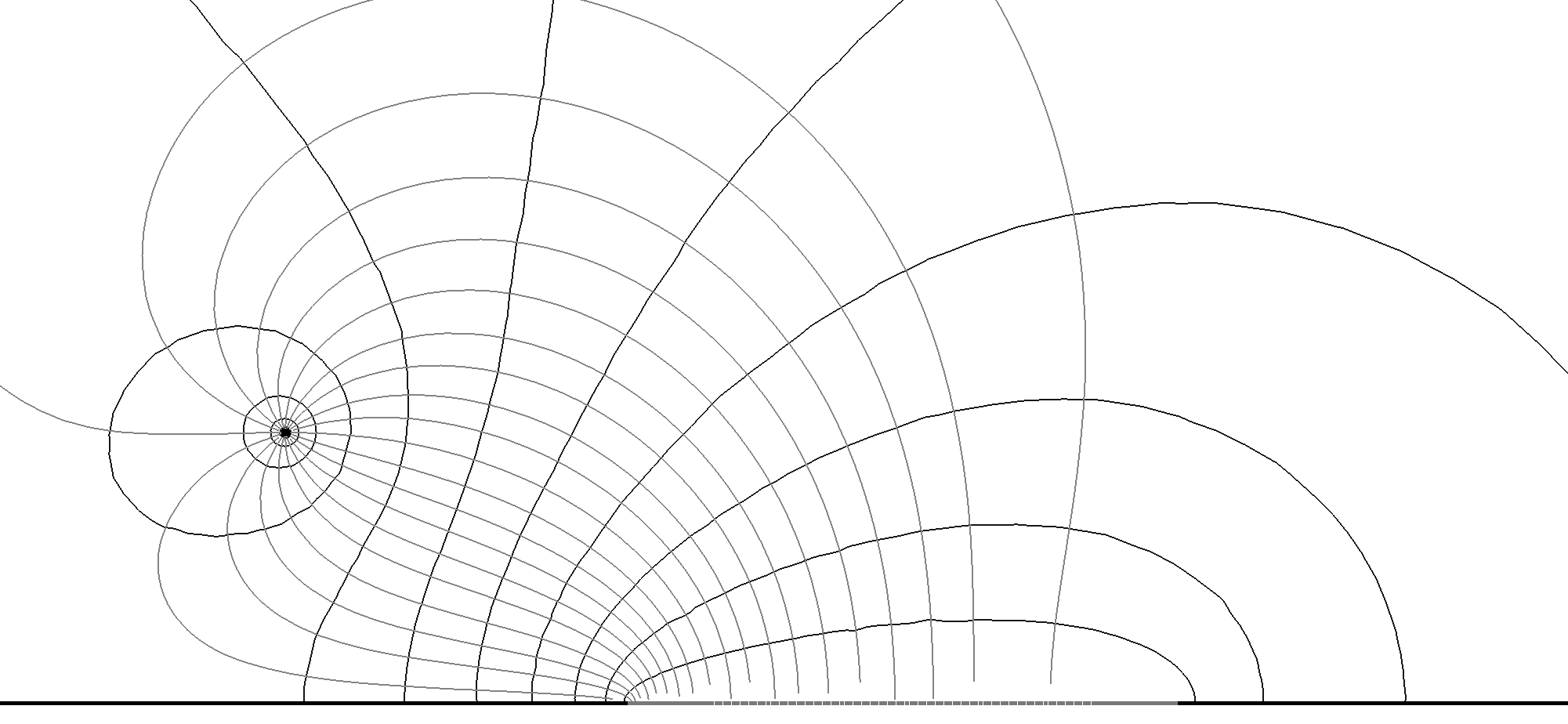}}
   \caption{Electric quantities in the device with open wire to wall-embedded collecting electrode arrangement discussed in Subsection~\ref{subsec:go} at a $3.6\,\mathrm{kV}$ applied voltage. The results shown here were obtained with the exponential diode condition.}\label{fig:go:electric}
   \centering
   \subfigure[\ Air velocity streamlines in a device region near the electrodes. The scale is the same as in Fig.~\ref{fig:go:el:1}.]{\includegraphics[width=.48\textwidth]{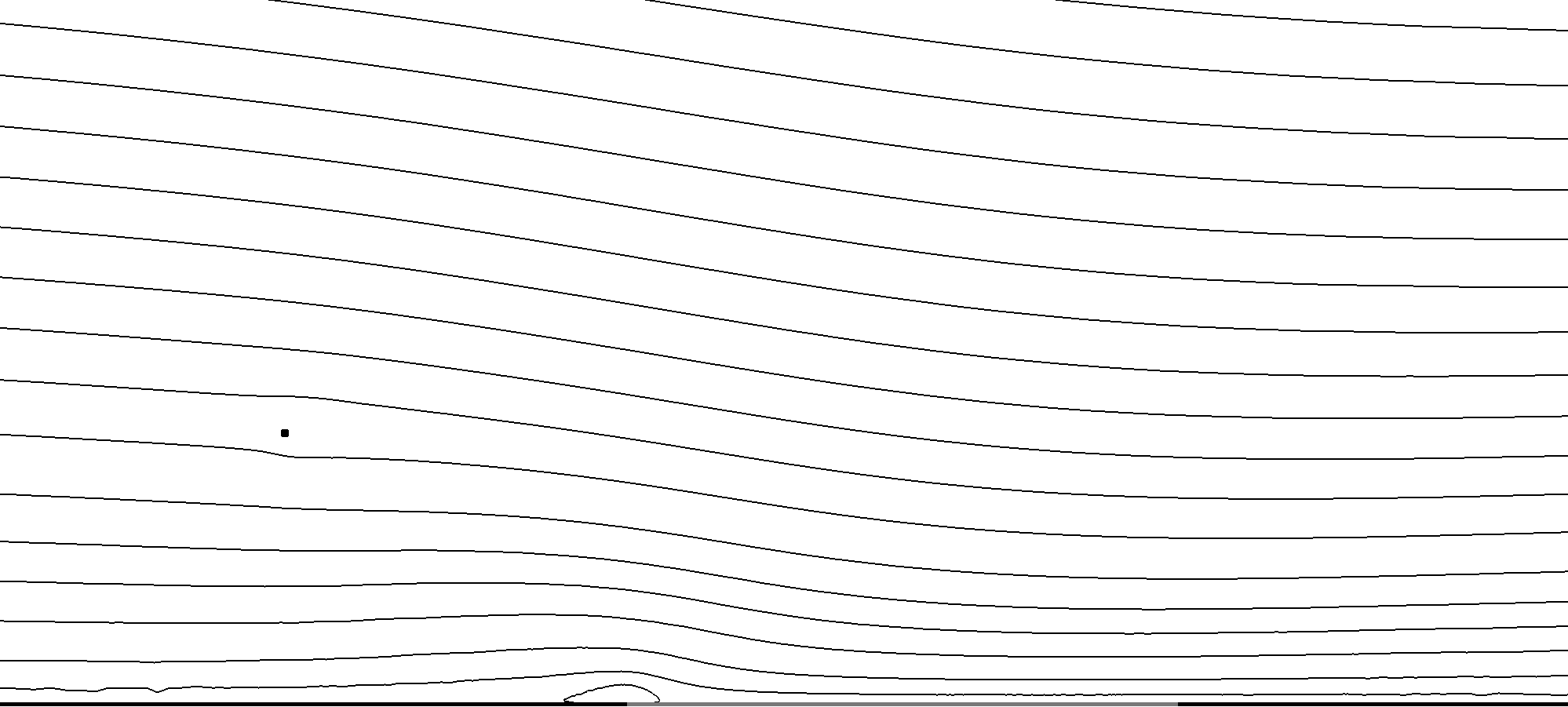}}
   \subfigure[\ Magnitude of air velocity ($\mathrm{m\ s^{-1}}$) in the whole computational domain. The ticks on the right show the length scale, each tick is $10\,\mathrm{mm}$.]
   {{\includegraphics[width=.48\textwidth]{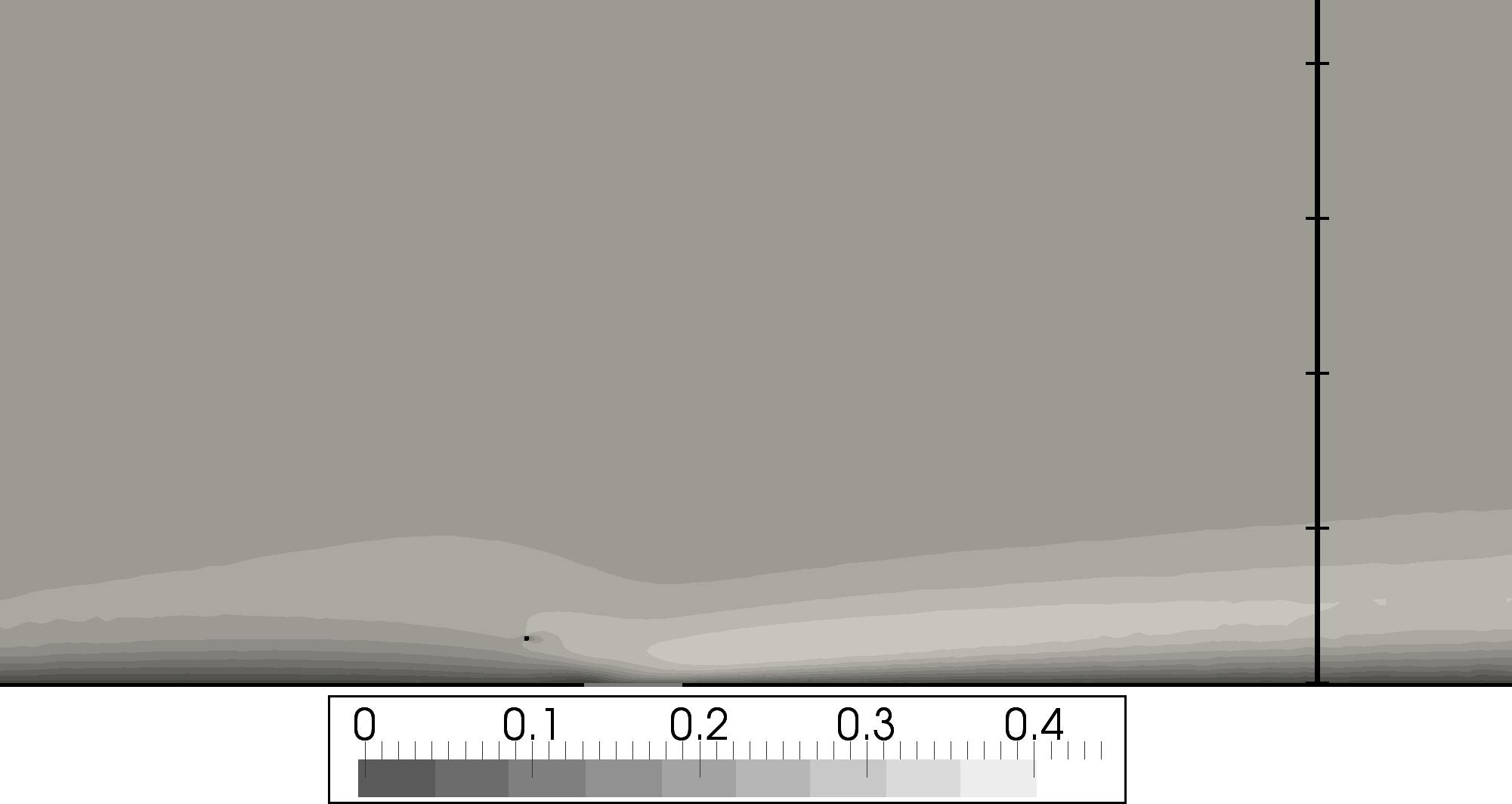}}}
   \caption{Air flow in the device with open wire to wall-embedded collecting electrode arrangement discussed in Subsection~\ref{subsec:go} at a $3.6\,\mathrm{kV}$ applied voltage. The results shown here were obtained with the exponential diode condition.} \label{fig:go:velo}
\end{figure}

\begin{figure}[t!] 
   \centering
   \includegraphics[width=.75\textwidth]{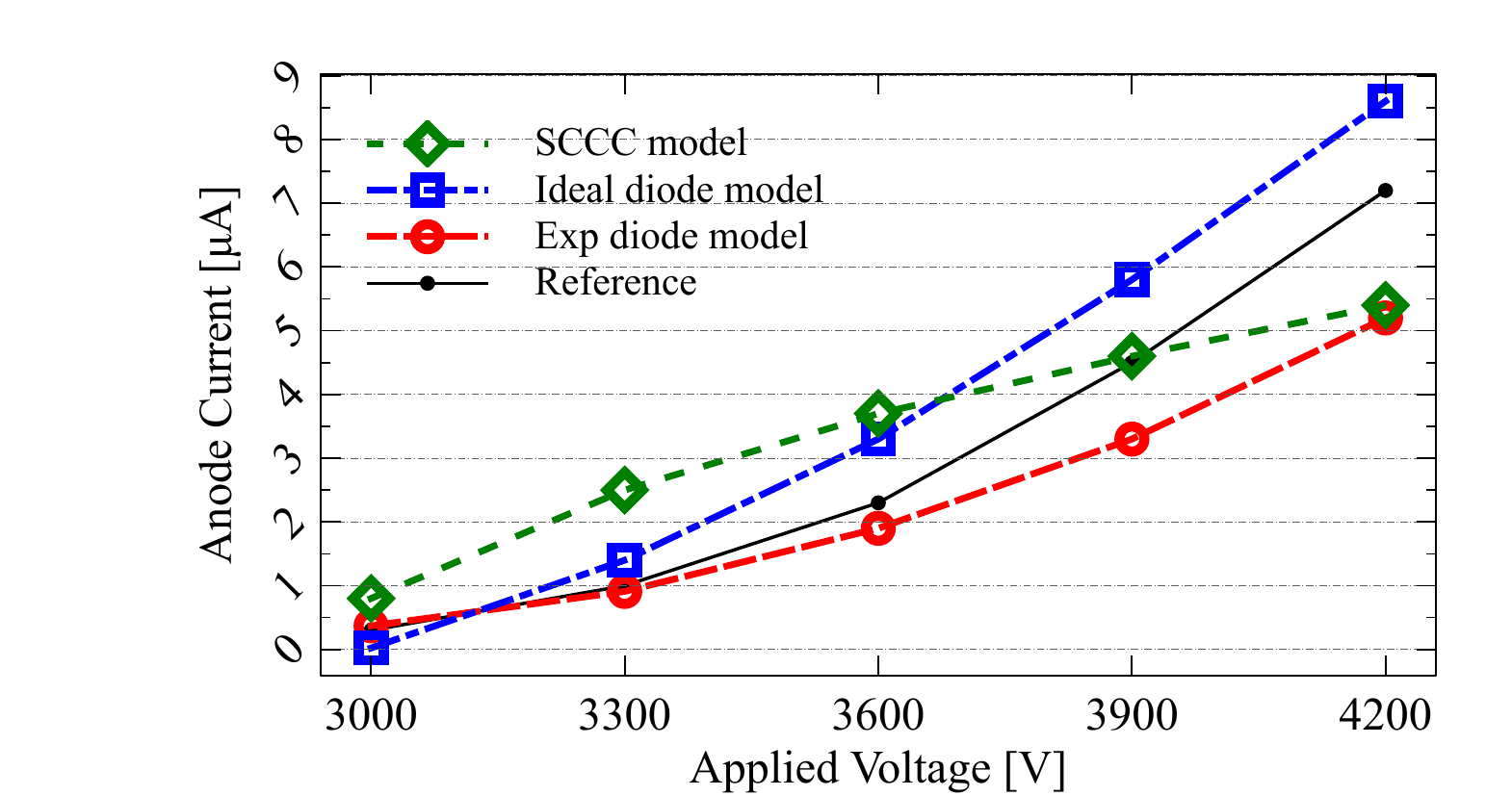}
   \caption{Anode current vs applied voltage in the device with open wire to wall-embedded collecting electrode arrangement discussed in Subsection~\ref{subsec:go}, computed applying the four boundary condition models presented in Section~\ref{sec:bcs}.}\label{fig:go:IV}
\end{figure}

In this section, we apply our numerical model to the wire-to-plate geometry studied in Refs.~\onlinecite{go-jap-07,go-ijhmt-08}. Figure~\ref{fig:Go:geometry} depicts the experimental setup: a flat, insulating plate $125\,\mathrm{mm}$ long and $50.8\,\mathrm{mm}$ wide, is placed in a laminar, $0.28\,\mathrm{m/s}$ air flow, parallel to the plate. In the plate is embedded a $6.35\,\mathrm{mm}$ long metal strip, its leading edge $55.25\,\mathrm{mm}$ away from the leading edge of the plate, acting as cathode contact. The $0.05\,\mathrm{mm}$ diameter wire acting as anode contact is placed $3.15\,\mathrm{mm}$ far from the plate, and $4\,\mathrm{mm}$ upstream of the cathode strip leading edge.

Figure~\ref{fig:go:electric} explains the working mechanism of the device. The main conductive channel is highlighted in Fig.~\ref{fig:go:el:1}: electric current flows mainly from the anode to the upstream part of the cathode, following the field lines depicted in Fig.~\ref{fig:go:el:2}. As shown in Fig.~\ref{fig:go:velo}, the generated EHD force and the wall reaction combine to reduce the thickness of the boundary layer in the region adjacent to the cathode strip.

Figure~\ref{fig:go:IV} compares the measured data with the results obtained with different models for the boundary condition. Only three models have been successfully used, since the \emph{uniform} model proved especially inappropriate in this very asymmetrical geometry: most of the charge injected from the anode side opposite to the cathode would stagnate, generating nonphysical solution as well as numerical misbehaving (due to the reformulation of Poisson's equation in Gummel's map algorithm).
The \emph{SCCC} model does not suffer of those issues, since no charge is injected from the low electric field side of the anode; nonetheless it fails to reproduce the correct, convex shape of the current-voltage curve, presenting an excessive shielding effect.

Both the \emph{ideal} and \emph{exponential diode} model provide better predictions, both qualitatively, with a convex IV curve, and quantitatively, with the maximum prediction error bounded under 33\% of the measured current. Additional accuracy could be obtained with a deeper research for the optimum parameters for both models, but this is beyond the scope of this work.

\subsection{Convergent duct with wire-to-plate electrode arrangement} \label{subsec:chang}
\begin{figure}[t]
\centering
 \includegraphics[width=.65\textwidth]{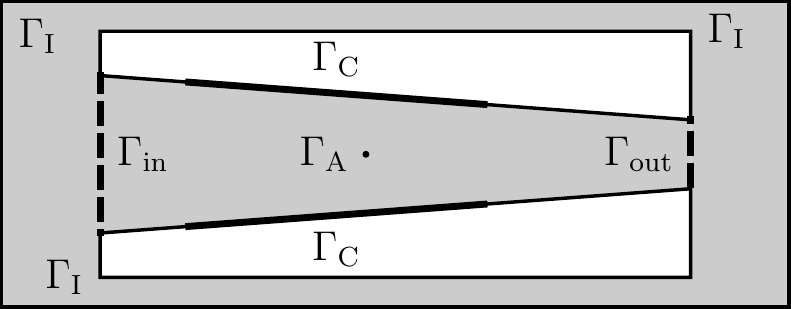}
 \caption{Schematic picture of the computational domain geometry for the device with convergent duct with wire-to-plate electrode arrangement discussed in Subsection~\ref{subsec:chang}.\label{fig:Ch:geometry}}
\end{figure}

\begin{figure}[t] 
   \centering
   \subfigure[\ Electric field lines (grey) and electric potential isolines (black).]{\includegraphics[width=.75\textwidth]{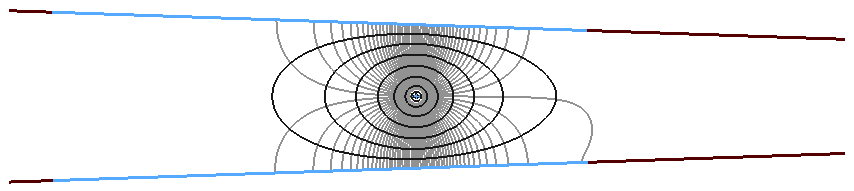}}
   \subfigure[\ Streamlines, sampled every $0.1\,\mathrm{s}$ from $t=2.6\,\mathrm{s}$ to $t=2.9\,\mathrm{s}$.]
   {\includegraphics[width=.75\textwidth]{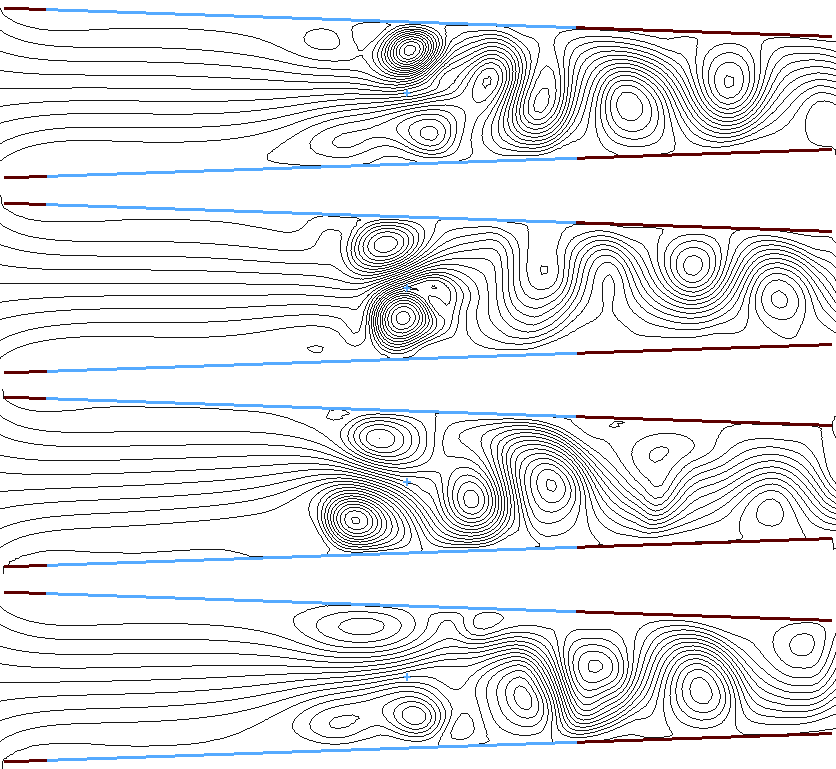}}
   \caption{Electric field (a) and air velocity (b) in the device with convergent duct with wire-to-plate electrode arrangement discussed in Subsection~\ref{subsec:chang} at an applied voltage of $9\,\mathrm{kV}$, these results were obtained with the exponential diode boundary condition.\label{fig:chang:space}}
\end{figure}

\begin{figure}[tb]
\centering
 \includegraphics[width=.75\textwidth]{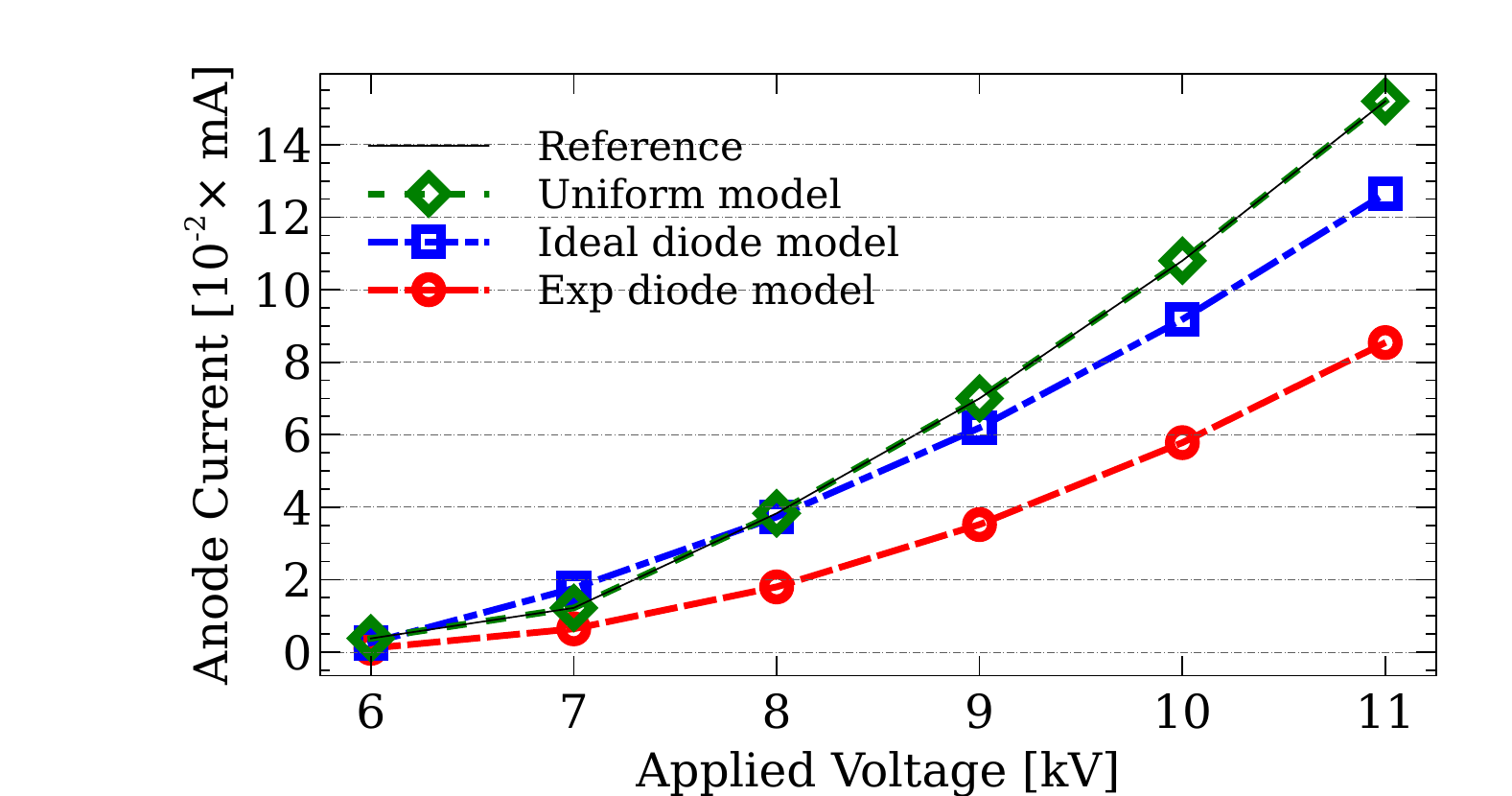}
 \caption{Anode current vs applied voltage in the device  with convergent duct with wire-to-plate electrode arrangement discussed in Subsection~\ref{subsec:chang}, computed applying three of the boundary condition models presented in  Section~\ref{sec:bcs}}\label{fig:chang:IV}
\end{figure}

\begin{figure}[bt] 
   \centering
   \subfigure[\ \emph{Exponential diode} model.]{\includegraphics[width=.65\textwidth]{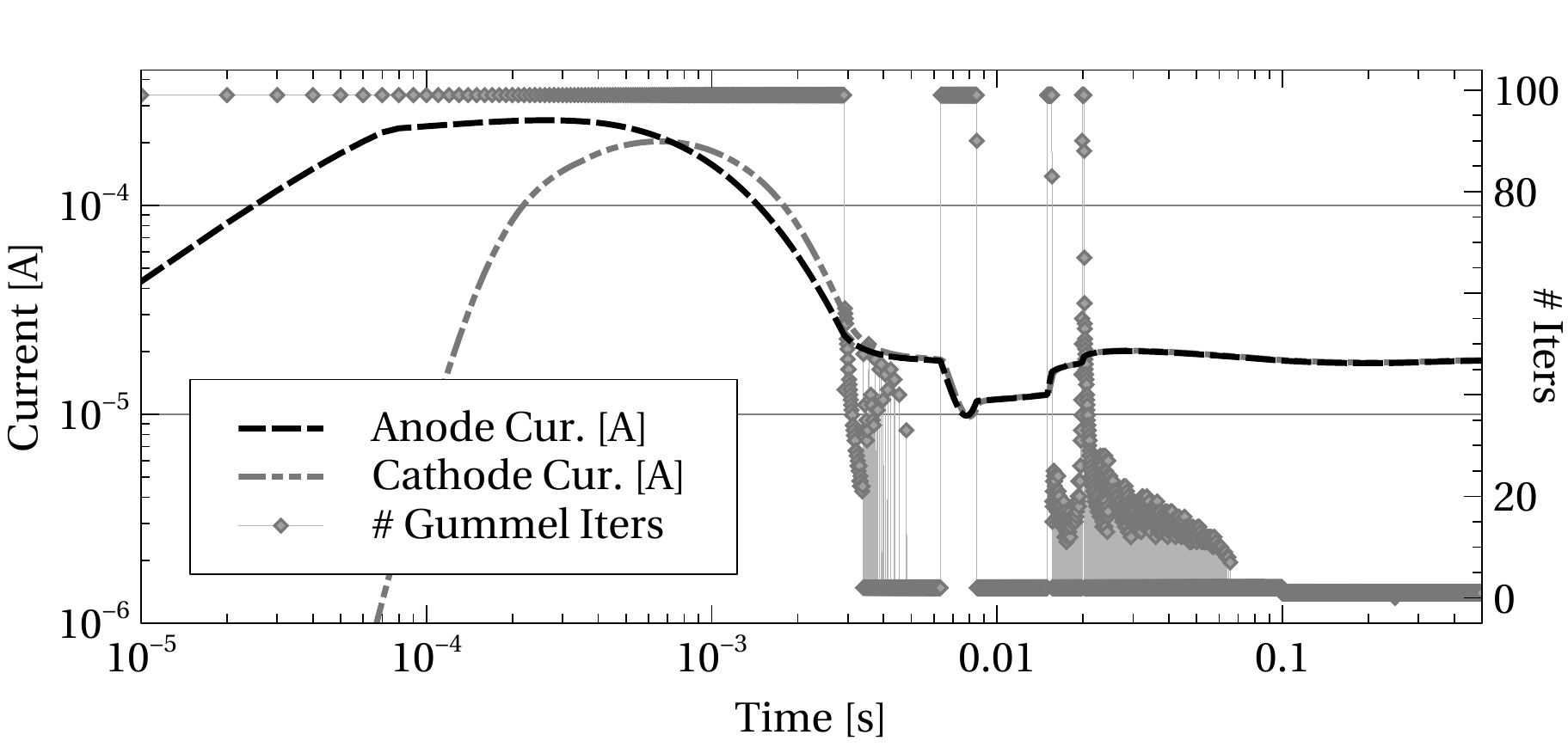}}
   \subfigure[\ \emph{Ideal diode} model.]{\includegraphics[width=.65\textwidth]{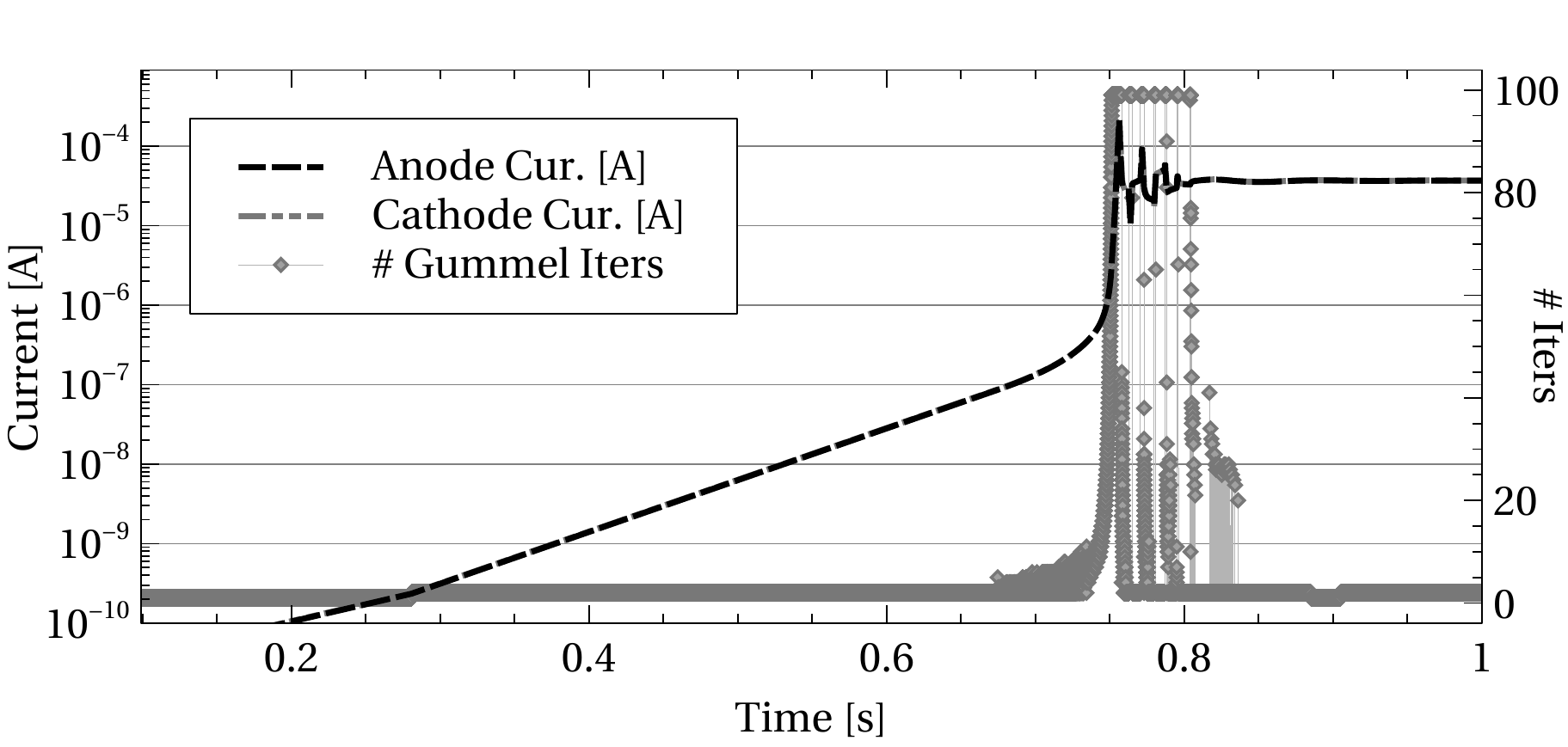}}
   \caption{Performance of the iterative algorithm in simulating the device with convergent duct with wire-to-plate electrode arrangement discussed in Subsection~\ref{subsec:chang} for an applied voltage of $8\,\mathrm{kV}$. The plots shows anode and cathode currents and the total number of iterations for the electric subsystem solution map $\mathcal{E}$ at each time step.} \label{fig:chang:algorithm}
\end{figure}

\begin{figure}[tb] 
   \centering
   \includegraphics[width=.75\textwidth]{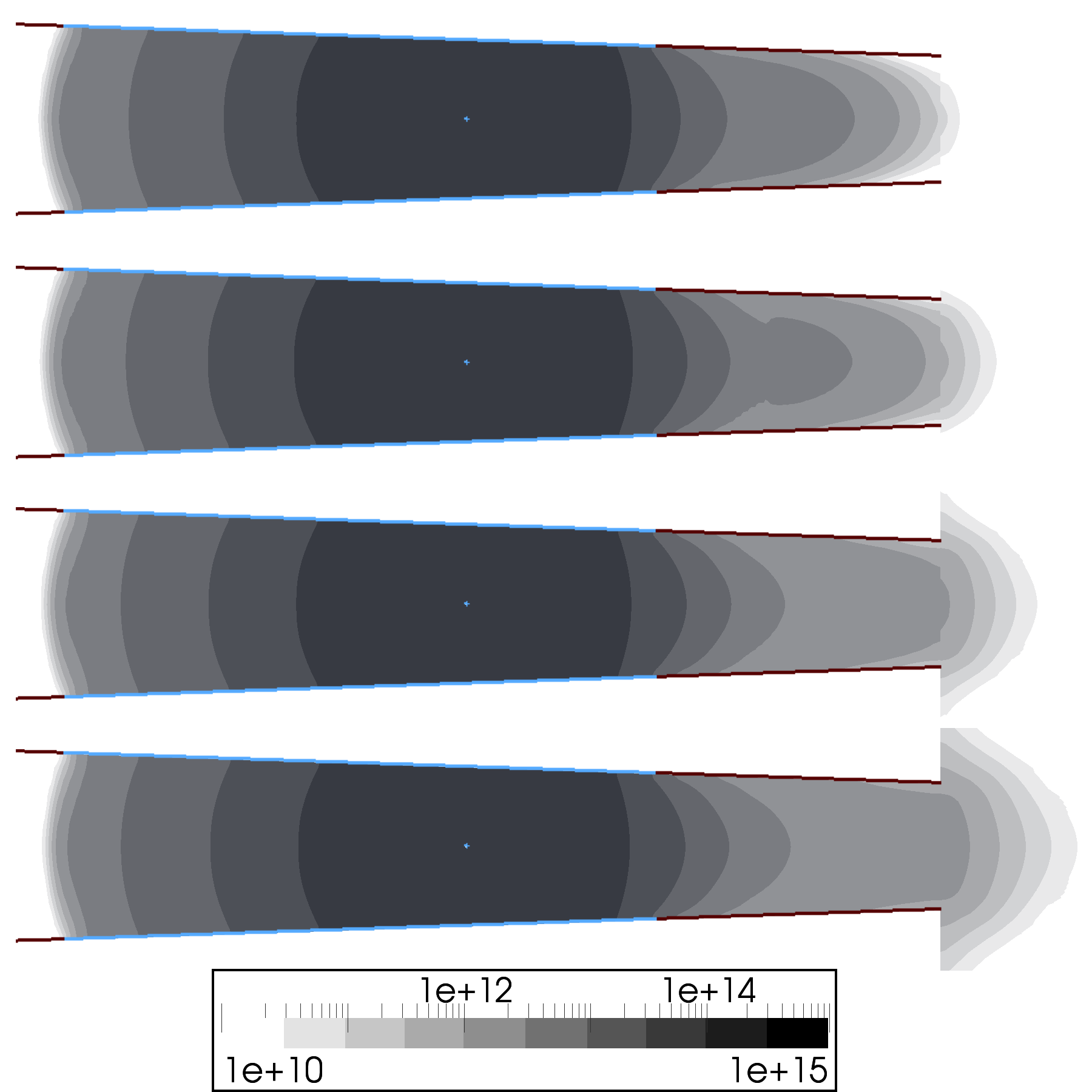}
   \caption{Charge distribution in the device with convergent duct with wire-to-plate electrode arrangement discussed in Subsection~\ref{subsec:chang}, sampled every $5\,\mathrm{ms}$ from $17\,\mathrm{ms}$ to $32\,\mathrm{ms}$ for an applied voltage of $8\,\mathrm{kV}$.} \label{fig:chang:space:charge}
\end{figure}

\begin{figure}[bt]
\centering
 \includegraphics[width=.75\textwidth]{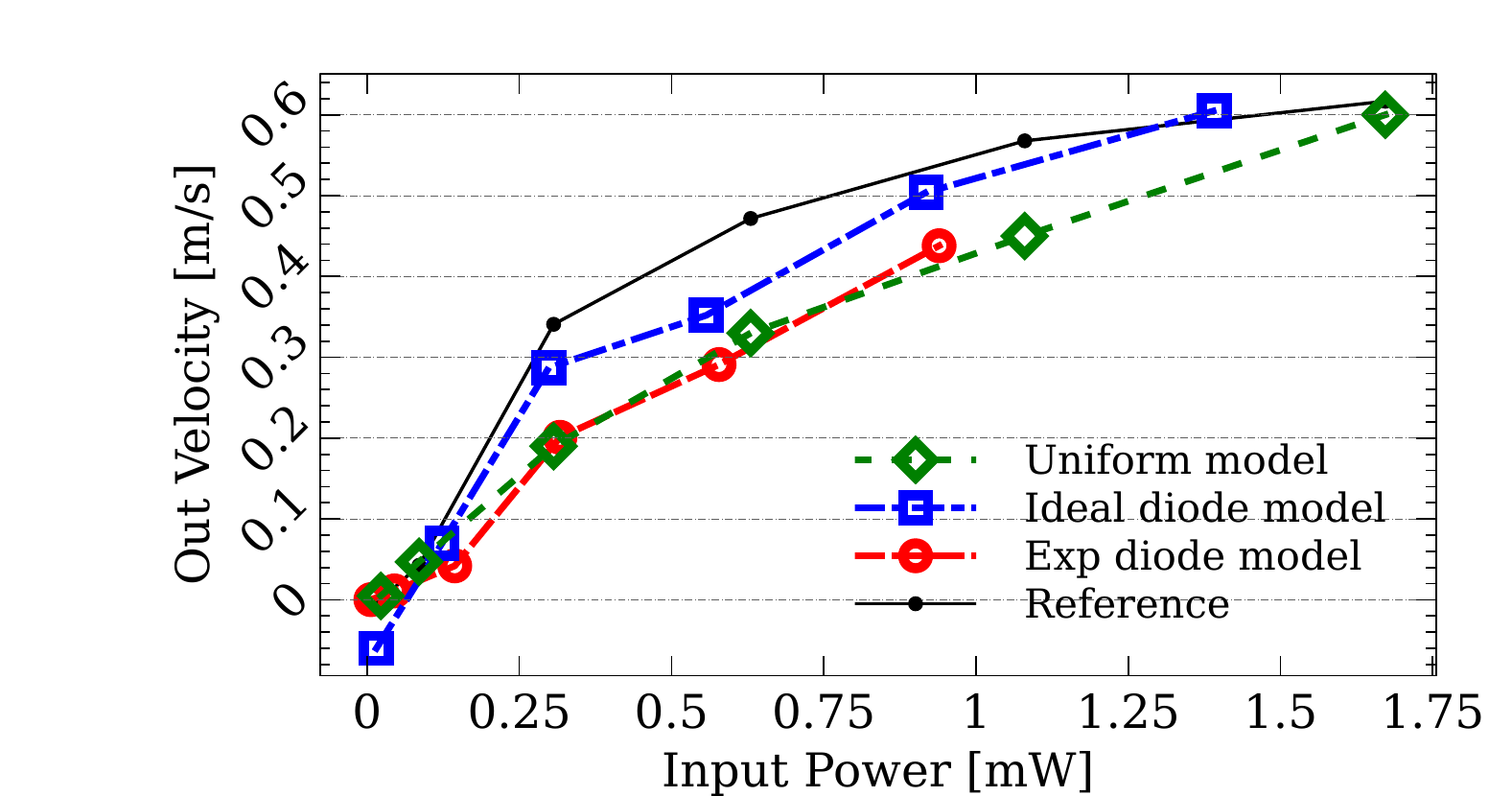}
 \caption{Steady state outlet velocity for the device with convergent duct with wire-to-plate electrode arrangement discussed in Subsection~\ref{subsec:chang}, computed applying three different boundary models from Section~\ref{sec:bcs}\label{fig:chang:PU}}
\end{figure}

In this section, we apply our numerical model to the device experimentally studied in Ref.~\onlinecite{chang-jpd-07}. The experimental setup is schematically represented in Fig.~\ref{fig:Ch:geometry}: a duct enclosed between two insulating non-parallel plates, $33\,\mathrm{mm}$ deep and $117\,\mathrm{mm}$ long, with the two openings $24$ and $12\,\mathrm{mm}$ wide, respectively. The wire acting as anode is placed $60\,\mathrm{mm}$ away from the smaller opening and has a diameter of $0.24\,\mathrm{mm}$. Two stripes of conductive material, acting as cathodes, are embedded on the non-parallel plates, ranging from $6\,\mathrm{mm}$ away from the wider opening to $36\,\mathrm{mm}$ away from the smaller one.

Figure~\ref{fig:chang:space} shows the basic working principle of the device. The electric field directed from the anode wire towards the cathode plates generates vortices, that are made non-symmetric by the reaction forces of the inclined walls. The non-symmetry produces a net air flow directed, for the particular electrodes arrangement at hand, from the wider cross-section to the smaller cross-section end. For high applied voltages, vortex shedding can be observed (see Fig.~\ref{fig:chang:space} again) and the flow becomes non-stationary (quasi-periodic).

Figure~\ref{fig:chang:IV} shows a comparison of the numerical simulation predictions for the anode current to applied voltage characteristics of the device. Simulations were performed with different injection models and compared to measurements from Ref.~\onlinecite{chang-jpd-07}. The \emph{uniform} model has been useful in this case, thanks to the symmetry of the domain, and matches by construction the experimental current values.
The currents predicted by the \emph{ideal diode} model appears to be in very good agreement with measurements both qualitatively and quantitatively, the relative error being consistently bounded under 17\% over a wide range of applied voltages.
The \emph{exponential diode} injection model also correctly captures the qualitative behavior of the IV curve, which is approximately parabolic in accordance to approximate analytic solution for totally axisymmetric geometries. The quantitative error with respect to the measurements is, as expected, higher. 

Such a loss in accuracy, though, is balanced by the better numerical performance. Figure~\ref{fig:chang:algorithm} compares the convergence history of the iterative method when the \emph{ideal diode} or the \emph{exponential diode} injection model is applied. The number of time steps required for the electric variables to reach a stationary regime is much higher in the case of the \emph{ideal diode} condition due to the requirement of a smaller under-relaxation coefficient that is needed to stabilize the method in this case. 
It is interesting to observe how the convergence over time of the current to its stationary value is non-monotonic. Indeed, a possibly high \emph{overshoot} in the current is usually observed, if the initial value of the cation density is low. In such situation, the anode contact electric field is initially much higher than at steady state, and thus more intense charge injection occurs. An additional abrupt change in the simulated current may occur, when the charge present in the device, due either to the initial value or the overshooting, is expelled from the channel as shown in Fig.~\ref{fig:chang:space:charge}; this abrupt change results in a variation of the anode charge density value, and leads to the need of a larger number of fixed point iterations. The above discussion shows that a careful choice of the initial condition is necessary in order to allow for a good performance of the numerical method.

Finally, Fig.~\ref{fig:chang:PU} shows a comparison of the experimental and predicted average velocities on the outlet section, plotted versus the total provided power at the electrical steady state $W = \I\Vapp$. The \emph{uniform} model only provides an approximation of scale of the total flow rate; on the other hand, it underestimates both the high increase in efficiency for smaller applied power and the drop in efficiency at higher power.
The \emph{ideal diode} model, on the contrary, provides a very good approximation for the efficiency of the device, due to the more realistic space distribution of the volume EHD force, even \emph{without} a-priori knowledge of the expected current. As already stated, this additional accuracy comes at the price of higher computational cost.
The \emph{real diode} approach, in the end, provides a flow rate curve quite similar to the one from the \emph{uniform} model, even if the points are biased towards the low-power region due to the underestimation of the currents. Moreover, the approach is not dependent on empirical data, since its parameters depend mainly on the electrode radius and could be estimated from similar cases.
This result is in our opinion a fair trade-off between the need of specific empirical data on currents of the \emph{uniform} model, and the excessive computational effort required by the \emph{ideal diode} model.

\FloatBarrier
\section{Industrial application example: an EHD cooled condensation radiator}

\begin{figure}[b!]
\centering
 \includegraphics[width=.7\textwidth]{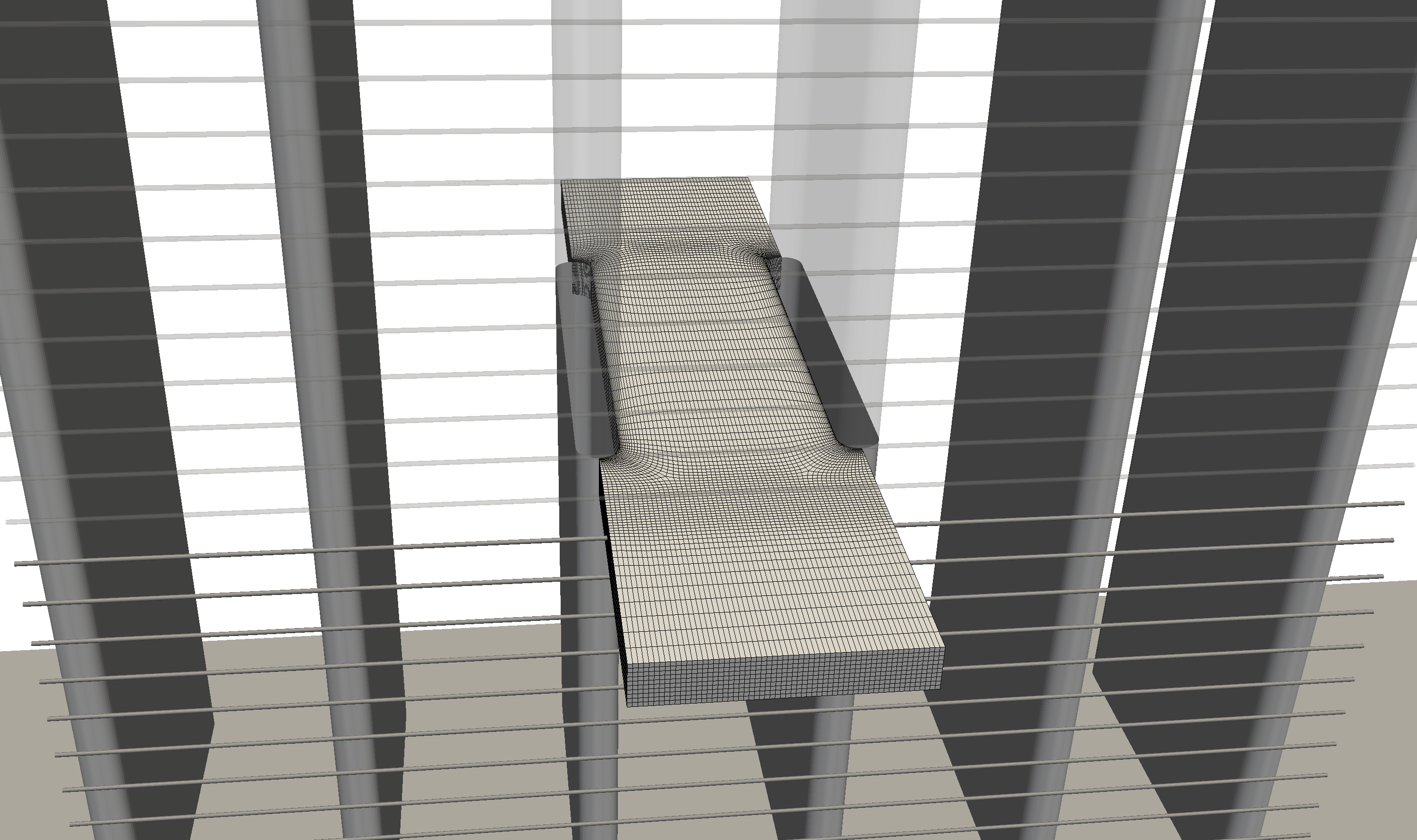}
 \caption{Geometry of the thermosyphon, with the mesh of wires acting as anode, and the basic periodic cell used as computational domain.\label{fig:thermo:geom}}
\end{figure}

After the model validation carried out in Section~\ref{sec:benchmark}, we present in this section an example of application of our simulation tool to the design of a cooling apparatus of potential impact in industrial application. In particular, we consider a combination of EHD forced air convection and a two-phase thermosyphon.

Two-phase cooling, and in particular two-phase thermosyphons, have been recognized in being beneficial for thermal management of electronics. The usage of pumpless systems together with dielectric fluids and high heat transfer coefficients demonstrated to be a perfect combination for cooling of electronics.

Thermosyphon condensers are commonly automotive type heat exchangers. 
This technology uses numerous multiport extruded tubes with capillary sized channels disposed in parallel and brazed to louvered air fins that meets the required compactness. The heat removal is obtained by means of a forced air stream of air over the condenser body usually imposed by a fan element. 
If fans represent a standard solution, drawbacks are commonly identified in the reliability (rotating mechanical parts), in the noise and in the occupied volume. 

An EHD cooled condenser can overcome the limits of a common fan system. For a given condenser size, the EHD cooler will increase the local air speed; for a given temperature of operation of the cooler, the EHD system can enable a global reduction of the system size with reduced noise levels. Last but not least, EHD can locally increase the condensing performances due to generated magnetic field enabling a reduction of the operating temperature of the electric and electronic devices.

The result we show in this section pertain to the simulation of a simplified model of EHD cooled thermosyphon, similar to the ones presented in Refs.~\onlinecite{agostini-11-INTELEC,agostini-12-IHPC,habert-12-IHPC,agostini-12-ITHERM}. Figure~\ref{fig:thermo:geom} depicts the geometry of the device, where the vertical tubes act as cathode and a mesh of thin wires acts as anode. The device is inherently modular, so that simulation is required only for the basic periodic cell, which is in evidence in Fig.~\ref{fig:thermo:geom}. On the horizontal boundary planes, periodic condition are imposed, while on the portions of the vertical boundary planes not intersecting the solid components, symmetry conditions are enforced. 
\FloatBarrier

\begin{figure}[b!]
\centering
 \includegraphics[width=.75\textwidth]{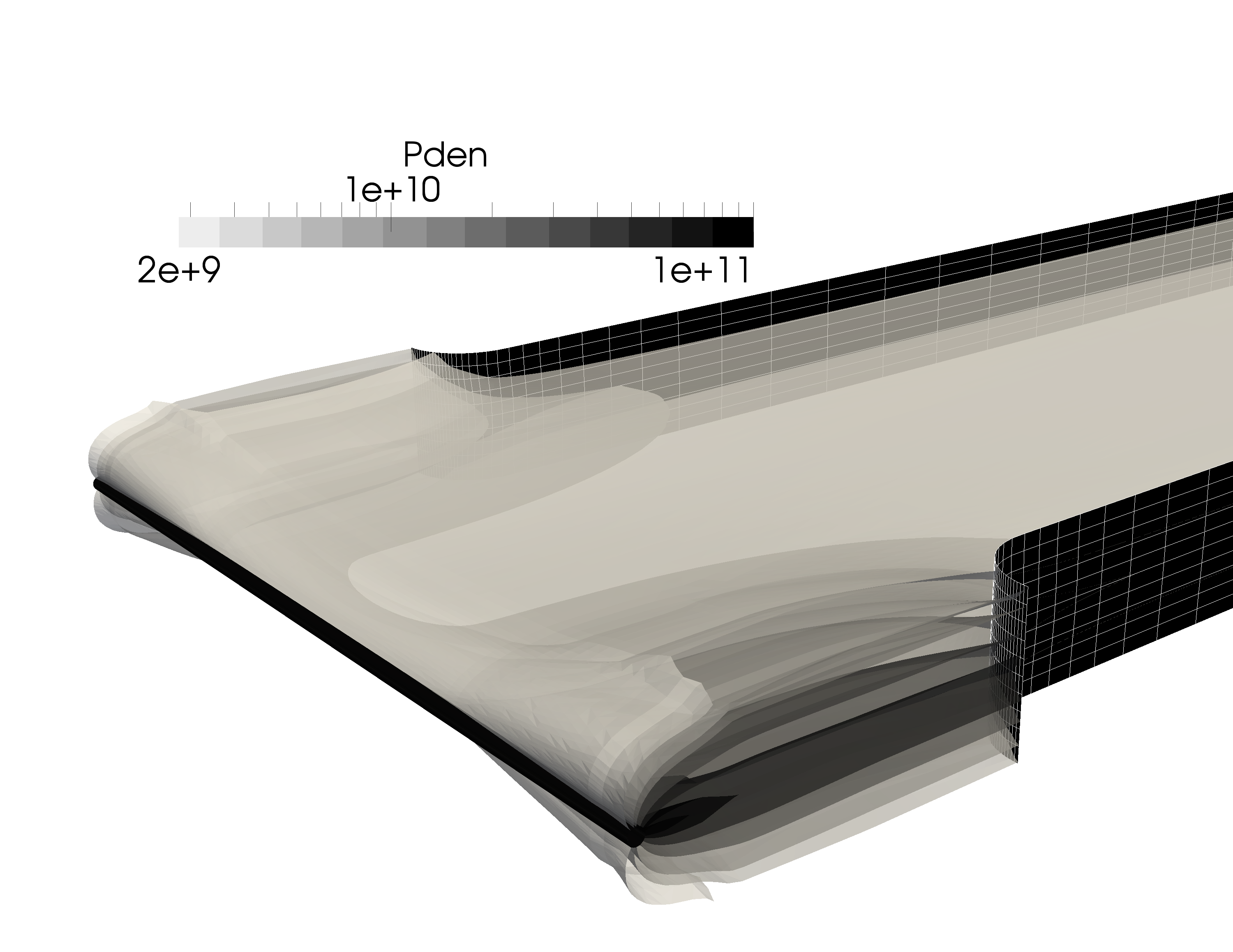}
 \caption{Cation number density ($\mathrm{m}^{-3}$) isosurfaces, for an applied voltage of $10\,\mathrm{kV}$.\label{fig:thermo:charge}}
 \centering
 \includegraphics[width=.75\textwidth]{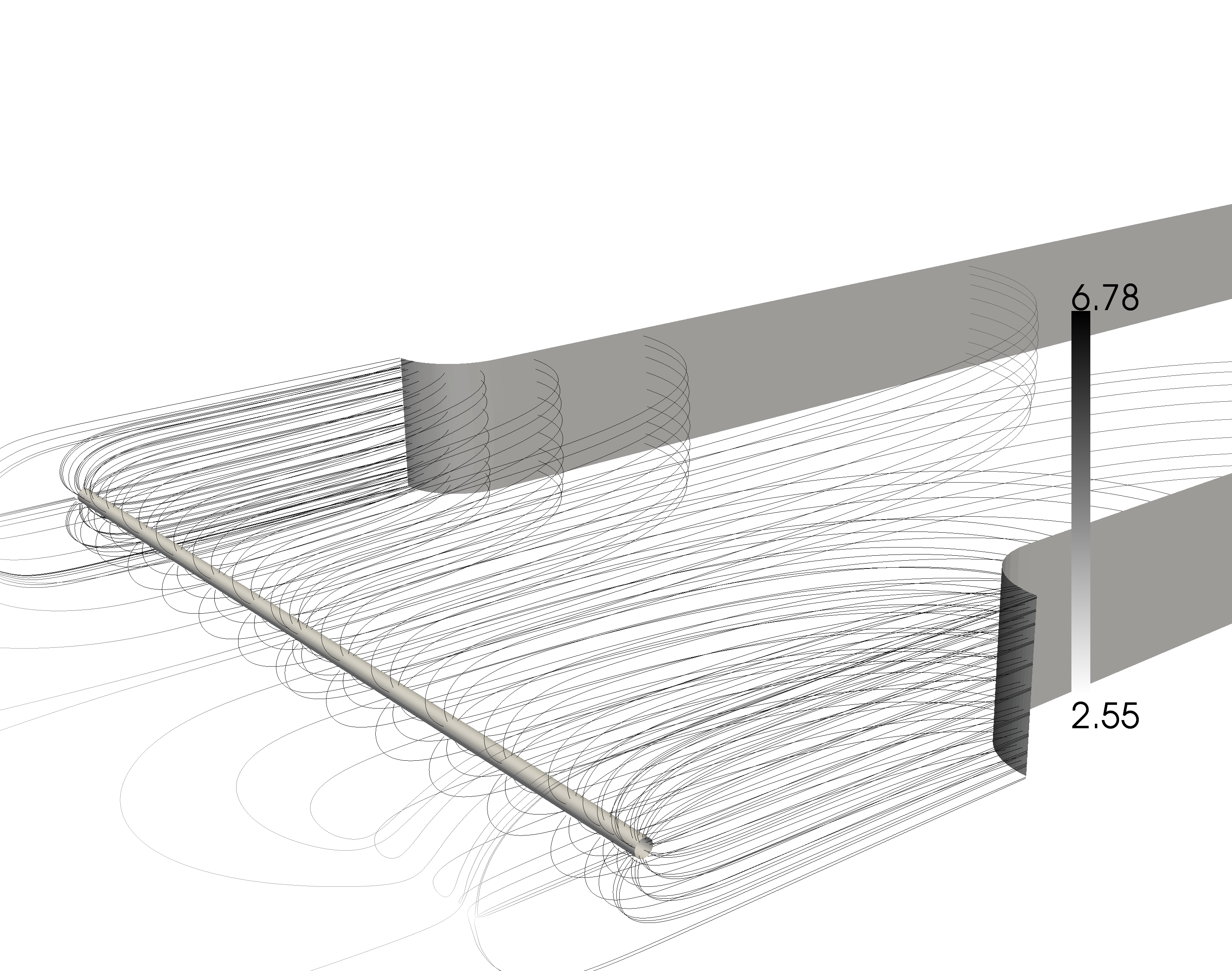}
 \caption{Electric field lines, color scale based on $\log_{10}$ of the magnitude of \E expressed in $\mathrm{Vm}^{-3}$, 
 for an applied voltage of $10\,\mathrm{kV}$.\label{fig:thermo:field}}
\end{figure}

Figure~\ref{fig:thermo:charge} shows the distribution of the cation density in the domain. The maximum density is located directly in front of the pipe, and a main conductive region is formed. Figure~\ref{fig:thermo:field} shows some electric field lines, which are also parallel to the EHD volume force, that triggers the fluid motion.

\FloatBarrier
\section{Conclusions}\label{sec:conclusions}
 
In this work, we studied the numerical approximation of the effects of electric discharge on ambient air flow. First, we proposed an algorithm to deal with the multiphysics mathematical model describing the system, by the coupling of the different and particular approaches already used in the fields of electronic device simulation and computational fluid dynamics.
Furthermore, we analyzed the particular phenomenon of corona discharge and proposed a phenomenological approach, which allows for the removal of the plasma subdomain and the electron density conservation equation from the computation. Four different models following this approach have been considered, discussed, and compared. The conclusion is that both the \emph{ideal} and \emph{exponential diode} models, proposed in this work, are able to reproduce the correct behavior of the corona discharge EHD system \emph{without} need of measured data for the electric current in the actual device at hand.
Finally, we showed how our models and algorithm can be effectively used in a relevant industrial application.


%

\end{document}